\documentclass{article}

 \usepackage{textcomp, mathtools}
 \usepackage{enumitem}
 \usepackage{tabularx}
\usepackage{array}

\newcolumntype{Y}{>{\raggedright\arraybackslash}X}
 \usepackage[preprint]{neurips_2025}

\usepackage[utf8]{inputenc} 
\usepackage[T1]{fontenc}    
\usepackage{hyperref}       
\usepackage{url}            
\usepackage{booktabs}       
\usepackage{amsfonts}       
\usepackage{nicefrac}       
\usepackage{microtype}      
\usepackage{xcolor}         
\usepackage{amsmath, amssymb, amsthm}
\usepackage{algorithm, algpseudocode}

\newtheorem{theorem}{Theorem}
\newtheorem{lemma}{Lemma}
\newtheorem{proposition}{Proposition}
\newtheorem{corollary}{Corollary}
\newtheorem{definition}{Definition}
\newtheorem{assumption}{Assumption}

\newtheorem{openproblem}{Open Problem}

\title{Field Codes for Distributed Coupling Samplers and Certified Empirical Transport}

\author{
  Hung PQ.~Mai\textsuperscript{1,2}
  \quad
  Duc Hai Nguyen\textsuperscript{1,3}
  \quad
  Luong Doan\textsuperscript{1}
  \quad
  Ngoc Vu\textsuperscript{1}
  \\
  Khanh Nguyen\textsuperscript{1,3}
  \quad
  Nhung Duong\textsuperscript{1}
  \quad
  Tuan Do\textsuperscript{1,4}
  \\[2mm]
  \textsuperscript{1}B0Labs, N2TP Technology
  \quad
  \textsuperscript{2}National Economics University, Vietnam
  \\
  \textsuperscript{3}Nanyang Technological University
  \quad
  \textsuperscript{4}Phenikaa University, Vietnam
  \\
  \texttt{Correspondence: tuando7758@gmail.com}
}

\begin{document}
 \algrenewcommand{\algorithmiccomment}[1]{\hfill \texttt{\# #1}}

\maketitle

\begin{abstract}
In this paper, we formulate three communication tasks for empirical optimal transport: distributed coupling sampling, cost-evaluable coupling output, and scalar value-certified sampling. Our main result is a field-code compiler: any communicated transport field approximating an optimal empirical Monge map to error $\eta$ can be completed by sparse target-cell residuals into an exact-marginal value-certified sampler with scalar certificate $W_1(\mu,\nu)\leq U\leq W_1(\mu,\nu)+2\Delta$, where $\Delta$ is the public target-partition diameter. The certificate accuracy is controlled by $\Delta$ alone. The field error $\eta$ controls residual communication under a cell-margin condition; without a margin, $\eta$ alone does not bound residuals. We instantiate the compiler via adaptive local-affine and tensor-product spline codes with $d(m+1)^db$ field bits in the spline case, plus residual lists charged separately. For lower bounds, exact Gap-Hamming embeddings prove certified output is hard, including a smooth cell-packing diffeomorphism family requiring $\Omega(\varepsilon^{-2d/(d+4)})$ communication for any cost-evaluable, cost-certified, or value-certified protocol. The same gadgets admit zero-communication samplers, formally separating the sampler and certificate-bearing output models. These results identify the transport field as the right communicated object whenever a field code is available, primarily as a residual-sparsity tool.

\end{abstract}

\section{Introduction}
\label{sec:introduction}

\subsection{Motivation}

Normalizing flows, flow matching, rectified flows, and OT-CFM methods learn objects that are fundamentally transport-based: maps, velocity fields, or couplings that move mass from a source distribution to a target distribution \citep{lipman2023flowmatching,liu2023rectifiedflow,tong2024otcfm,tong2024sf2m,guo2025vrfm}. In standard centralized training, the learner can access both source and target samples and can construct pairs, costs, or velocity targets directly. This assumption breaks in distributed settings. If the source samples and target samples are held by different parties, then the transport object is no longer merely a computational output; it becomes an information-theoretic object that must be communicated. The basic question is therefore: \emph{what must be transmitted in order to construct a coupling, not just estimate its cost?}

This distinction is important for flow-based learning. A scalar estimate of a Wasserstein distance may certify that two empirical distributions are close, but it does not tell the source party which target atom should be paired with a source atom, nor does it provide the velocity target \(y-x\) used by flow-matching-style objectives. Coupling output is therefore a stronger and more operational task than distance estimation. The paper studies this task directly: under a bit budget, which representation should be communicated---target counts, support prototypes, an explicit plan, or a transport field?

Let \(\|\cdot\|_2\) denote the Euclidean norm, write \(\delta_z\) for the Dirac mass at a point \(z\), and let \(\Pi(\mu,\nu)\) denote the set of couplings of two probability measures \(\mu\) and \(\nu\). One party observes the source sample set \(X=\{x_1,\ldots,x_n\}\subset[0,1]^d\), while the other observes the target sample set \(Y=\{y_1,\ldots,y_n\}\subset[0,1]^d\). These define the empirical measures \(\mu=\frac1n\sum_{i=1}^n\delta_{x_i}\) and \(\nu=\frac1n\sum_{i=1}^n\delta_{y_i}\). The distributed task is to communicate enough information to output a coupling between \(\mu\) and \(\nu\). The quality measure is the Wasserstein-1 transport cost
\[
W_1(\mu,\nu)
=
\inf_{\pi\in\Pi(\mu,\nu)}
\int \|x-y\|_2\,d\pi(x,y),
\]
and the communication measure is the number of transmitted bits.

The output model matters because different ways of exposing a coupling reveal different amounts of information. A protocol that samples matched pairs, a protocol that reveals an explicit public transport plan, a protocol whose coupling cost is evaluable by one party, and a sampler equipped with a scalar value certificate are distinct communication tasks. This paper separates these notions and asks which transport representation is sufficient for each task.

The contribution of this paper is to answer this question through a direct coupling-output formulation for empirical OT. We show that a communicated transport field, combined with sparse residual target-cell counts, can be compiled into an exact-marginal distributed sampler with a scalar value certificate. We also show that stronger certificate-bearing outputs, namely cost-evaluable, cost-certified, and value-certified protocols, remain communication-hard even on smooth map-induced instances. Thus, field communication is not only a modeling choice inherited from flow methods; it is also a natural information-theoretic object for making distributed coupling output possible. We instantiate this principle through compiler-style upper bounds, smooth diffeomorphic lower-bound families, and empirical comparisons between field-based summaries, target-count summaries, and support-prototype summaries under explicit bit budgets.


\subsection{Research Questions}
\label{subsec:research-questions}

The paper formalizes four research questions. RQ1 asks for the right two-party formulation for outputting an approximate coupling, as opposed to estimating an OT cost. RQ2 asks which communicated object gives a constructive upper bound in smooth map-induced regimes, among target counts, support prototypes, and transport fields. RQ3 asks whether smooth map-induced OT instances are communication-hard. RQ4 asks whether field-based summaries are empirically useful under explicit bit budgets on synthetic and natural benchmarks.

\subsection{Contributions and Evidence}
\label{subsec:contributions-evidence}

Definitions~\ref{def:empirical-ot-output-model}--\ref{def:certificate-bearing-output} settle the formulation question RQ1 by distinguishing three output strengths: distributed samplers, cost-evaluable couplings, and scalar value-certified samplers. Theorem~\ref{thm:field-code-compiler} is the field-code compiler: any communicated field code with error $\eta$ is completed into an exact-marginal sampler equipped with a scalar value certificate of additive error $2\Delta$, where $\Delta$ is the diameter of a public target partition. The field error $\eta$ does not enter the certificate bound; it is the quantity relevant to residual communication, but does not by itself bound the residual lists without a margin or sparsity assumption. Smooth constructive instantiations of the compiler are Theorem~\ref{thm:adaptive-local-affine-code}, an adaptive local-affine code controlled by local curvature with explicit $\varepsilon$-rate in Corollary~\ref{cor:adaptive-local-affine-rate}, and Theorem~\ref{thm:tensor-product-spline-code}, a dimension-explicit tensor-product spline code with explicit value-certified field-bit rate in Corollary~\ref{cor:spline-field-bit-rate}; together these answer RQ2. Certificate-bearing communication lower bounds for RQ3 come from Theorem~\ref{thm:bounded-support-separated-gap-hamming}, a bounded-support separated Gap-Hamming embedding, and Theorem~\ref{thm:smooth-cell-packing-diffeomorphism}, a smooth cell-packing diffeomorphism embedding with exponent $2d/(d+4)$ that binds against cost-evaluable, cost-certified, and value-certified output alike. Proposition~\ref{prop:zero-communication-sampler-separation} supplies model separation by exhibiting zero-communication distributed samplers for the same Gap-Hamming gadgets, scoping the lower bounds to the certificate-bearing models. The empirical study in Section~\ref{sec:experiments} addresses RQ4 by evaluating support summaries, grids, affine fields, local-affine fields, and spline fields under matched bit accounting on smooth synthetic tasks, 5D scale-ups, MNIST to USPS in shared PCA-5, and DOTmark ClassicImages.

The central message is theorem-first: in smooth empirical OT, a low-complexity transport field is the communication object that makes coupling output possible. Residual target-mass mismatch is handled afterward by sparse cell corrections, and the experiments study that design principle under explicit bit budgets.

\section{Related Work}

This work connects communication complexity, optimal transport, and flow-based generative modeling. We use Yao's two-party communication framework~\citep{yao1979distributive}, with lower bounds based on hard predicates such as Gap-Hamming~\citep{sherstov2012ghd} and Set-Disjointness~\citep{hastad2007setdisjointness}; distributed approximate matching lower bounds~\citep{huang2015matching} are conceptually related because empirical OT on separated supports behaves like weighted bipartite matching. Prior EMD sketching and streaming work studies compression of transportation distances~\citep{andoni2008emdhighdim,andoni2009emdsketch,chen2022streamingemd,andoni2016w2sketch}, while Sinkhorn, sliced/projection-robust OT, decentralized barycenters, and federated transport focus on scalable computation or networked optimization~\citep{cuturi2013sinkhorn,altschuler2017sinkhorn,peyre2019computationalot,lin2021projectionrobust,nadjahi2021fastsw,nguyen2024rpsw,staib2017streamingbarycenters,dvurechensky2018decentralize,uribe2018distributedbarycenters,cisnerosvelarde2020displacement,krishnan2025distributedot,kulcsar2025federatedsinkhorn}. In parallel, Flow Matching, Rectified Flow, OT-CFM, simulation-free bridge methods, and structured transport-map models learn maps or velocity fields related to couplings~\citep{lipman2023flowmatching,liu2023rectifiedflow,tong2024otcfm,tong2024sf2m,guo2025vrfm,cuturi2023sparsemaps,sidheekh2022vqflows}. Unlike these lines of work, we study the communication needed to output a coupling with exact empirical marginals and certificates, and we identify transport fields plus residual counts as the central communication object.

\section{Problem Formulation and Assumptions}
\label{sec:problem-formulation}

\subsection{Coupling-Output Communication}
\label{subsec:coupling-output-communication}

\begin{definition}[Distributed coupling sampler]
\label{def:empirical-ot-output-model}
\label{def:distributed-coupling-sampler}
Let $C(x,y)=\|x-y\|_2$. Let Alice denote the source party holding $X=\{x_1,\ldots,x_n\}$, and let Bob denote the target party holding $Y=\{y_1,\ldots,y_n\}$. A distributed coupling sampler is a public-coin two-party protocol in which Alice and Bob communicate a transcript $\mathsf{T}$ and then use public randomness together with their local inputs to sample a pair $(x_i,y_j)$. The protocol is valid if the induced law $\Pi_{\mathsf{T}}$ on $X\times Y$ has marginals exactly $\mu$ and $\nu$ almost surely. Its expected cost is $\mathbb{E}_{\mathrm{prot}}\langle \Pi_{\mathsf{T}},C\rangle=\mathbb{E}_{\mathrm{prot}}\sum_{i,j}\Pi_{\mathsf{T}}(i,j)\|x_i-y_j\|_2$. An additive-$\varepsilon$ expected-cost protocol satisfies $\mathbb{E}_{\mathrm{prot}}\langle \Pi_{\mathsf{T}},C\rangle \leq W_1(\mu,\nu)+\varepsilon$ on every instance in the class. The transcript length is the communication cost.
\end{definition}

This sampler model is stronger than distance sketching because it must instantiate matched mass. It is weaker than a public explicit plan, because Bob may use his private target samples during the final sampling step.

\begin{definition}[Cost-evaluable, cost-certified, and value-certified coupling output]
\label{def:certificate-bearing-output}
\label{def:cost-evaluable-cost-certified-value-certified}
Three output strengths beyond Definition~\ref{def:distributed-coupling-sampler} are distinguished. In all three, success probability is denoted $p\in(0,1]$.

A cost-evaluable coupling protocol outputs a representation of the induced coupling distribution $\Pi_{\mathsf T}$, not merely one sampled pair, such that Alice can compute the expected transportation cost $\langle \Pi_{\mathsf T},C\rangle$ from her final view: her input, public randomness, transcript, and the output representation. Any representation needed for this computation is either computable from that final view or must be communicated and counted in the protocol cost. An additive-$\varepsilon$ cost-evaluable protocol satisfies $\Pr\!\left[\Pi_{\mathsf T}\text{ is a valid coupling and }\langle \Pi_{\mathsf T},C\rangle\leq W_1(\mu,\nu)+\varepsilon\right]\geq p$. Since validity implies $\langle \Pi_{\mathsf T},C\rangle\geq W_1(\mu,\nu)$, the evaluable cost is itself an additive-$\varepsilon$ estimator on the success event.

A protocol is cost-certified if it is cost-evaluable and additionally outputs a scalar $\widehat W$ that, on the success event, satisfies $W_1(\mu,\nu)\leq \widehat W\leq W_1(\mu,\nu)+\varepsilon$.

A protocol is a scalar value-certified sampler if it is a valid distributed sampler in the sense of Definition~\ref{def:distributed-coupling-sampler} and additionally produces a scalar $U$ available to Alice from her final view, namely input, public randomness, transcript, and any additional output representation, such that, with probability at least $p$, $W_1(\mu,\nu)\leq U\leq W_1(\mu,\nu)+\varepsilon$. Any communication required so that $U$ becomes computable from Alice's final view is part of the protocol cost. A value-certified sampler need not be cost-evaluable: Alice may not be able to compute the realized expected sampler cost $\langle \Pi_{\mathsf T},C\rangle$ from her final view, but she does have a scalar $W_1$ upper bound.
\end{definition}

The three notions are distinct. Cost-evaluable, cost-certified, and value-certified protocols all imply additive-$\varepsilon$ estimation of $W_1(\mu,\nu)$. A bare distributed sampler does not: a public sampler may generate the correct pair structure without revealing the numerical transport cost.

\subsection{Standing Assumptions}
\label{subsec:standing-assumptions}

The theory uses at most five assumptions. Assumptions~\ref{assump:bounded-euclidean-empirical-ot}--\ref{assump:public-target-partition} define the common coupling-output model. Assumptions~\ref{assump:smooth-empirical-monge-realization}--\ref{assump:field-complexity-condition} are invoked only by the smooth constructive instantiations.

\begin{assumption}[Bounded Euclidean empirical OT]
\label{assump:bounded-euclidean-empirical-ot}
The samples lie in $[0,1]^d$, the transport cost is $C(x,y)=\|x-y\|_2$, and the empirical measures are uniform: $\mu=\frac1n\sum_{i=1}^n\delta_{x_i}$ and $\nu=\frac1n\sum_{i=1}^n\delta_{y_i}$.
\end{assumption}

\begin{assumption}[Public communication model]
\label{assump:public-communication-model}
The parties use public randomness and shared public quantizers. A sampler is valid only if the returned randomized coupling has marginals exactly $\mu$ and $\nu$ almost surely. Lower bounds specify when the stronger cost-evaluable, cost-certified, or scalar value-certified output model is required.
\end{assumption}

\begin{assumption}[Public target partition]
\label{assump:public-target-partition}
For the constructive protocols, both parties know a target partition $\{C_j\}_{j=1}^M$ with representatives $z_j\in C_j$ and Euclidean cell diameter at most $\Delta$.
\end{assumption}

\begin{assumption}[Smooth empirical Monge realization]
\label{assump:smooth-empirical-monge-realization}
There is an optimal empirical permutation $\sigma$ and a map $T:[0,1]^d\to[0,1]^d$ such that $T(x_i)=y_{\sigma(i)}$ for all $i\in[n]$.
\end{assumption}

\begin{assumption}[One field-complexity condition]
\label{assump:field-complexity-condition}
In each constructive theorem, a preliminary field-code stage produces a field satisfying exactly one of the following representation conditions:
\begin{enumerate}
    \item \textbf{Field approximation}: a communicated field $F_\theta$ satisfies $\max_i\|F_\theta(x_i)-T(x_i)\|_2\leq \eta$.

    \item \textbf{Adaptive local-affine smoothness}: $T$ is $C^{1,1}$ and each active dyadic leaf $Q$ satisfies the Lipschitz-gradient condition $\operatorname{Lip}(DT;Q)\operatorname{diam}(Q)^2\leq 8\tau$, where $\operatorname{Lip}(DT;Q)$ is the operator-norm Lipschitz constant of $DT$ on $Q$.

    \item \textbf{Tensor-product spline smoothness}: every component $T_k$ is $C^2$ and $\|\partial_{jj}T_k\|_\infty\leq H$ for all $j,k\in[d]$.
\end{enumerate}
\end{assumption}

These assumptions are appropriate for the paper's object of study for three reasons. First, normalizing flows and continuous transport models represent maps or velocity fields, so Assumptions~\ref{assump:smooth-empirical-monge-realization}--\ref{assump:field-complexity-condition} are the natural communication analogues of a flow parameterization. Second, public quantization in Assumptions~\ref{assump:public-communication-model}--\ref{assump:public-target-partition} is the standard shared-coordinate structure needed for bit-counted protocols with exact empirical marginals. Third, Theorem~\ref{thm:smooth-cell-packing-diffeomorphism} proves certified-output hardness inside a uniformly smooth diffeomorphic subclass, establishing that smoothness is also where the lower bounds bind.

\section{Theoretical Results}
\label{sec:theoretical-results}
This section gives the main theoretical results of the paper. We first present a shared-grid baseline that communicates target cell counts, then introduce the field-code compiler that turns a communicated transport field and sparse residual counts into an exact-marginal value-certified sampler. We instantiate the compiler with adaptive local-affine and tensor-product spline field codes, and then prove certified communication lower bounds together with a sampler/certificate separation. Full proofs of all theoretical results are provided in Appendix~\ref{app:full-proofs}; the main text states the results and highlights their communication consequences.

\subsection{A Grid Coupling Baseline}
\label{subsec:grid-coupling-baseline}


\begin{proposition}[Shared-grid coupling]
\label{prop:shared-grid-coupling}
Let $\{C_j\}_{j=1}^M$ have diameter at most $\Delta$. Bob sends the counts $m_j=\#\{i:y_i\in C_j\}$ using $M\lceil\log_2(n+1)\rceil$ bits. Alice forms $\tilde\nu=\sum_{j=1}^M\frac{m_j}{n}\delta_{z_j}$. An optimal coupling from $\mu$ to $\tilde\nu$, followed by Bob's within-cell lift from $z_j$ to the actual points in $C_j$, gives a valid randomized coupling between $\mu$ and $\nu$ with expected cost at most $W_1(\mu,\nu)+2\Delta$.
\end{proposition}


The main upper bound improves this by using a communicated transport field before residual counts are sent.

\subsection{Field-Preconditioned Residual Coupling}
\label{subsec:field-preconditioned-residual-coupling}

\begin{theorem}[Field-code-to-sampler-and-certificate compiler]
\label{thm:field-code-compiler}
Let $\mu,\nu$ be empirical measures in $[0,1]^d$. Fix a public target partition of diameter at most $\Delta$. Suppose a preliminary communication stage of $B_{\mathrm{field}}$ bits leaves Alice able to evaluate a public-family field $F_\theta$ satisfying, for some optimal empirical permutation $\sigma$, $\|F_\theta(x_i)-y_{\sigma(i)}\|_2\leq\eta$ for all $i\in[n]$. Then there is a public-coin two-round compiler that outputs a scalar value-certified sampler in the sense of Definition~\ref{def:certificate-bearing-output}: a distributed sampler with exact marginals $\mu,\nu$ almost surely and
\[
\mathbb{E}_{\mathrm{prot}}\langle \Pi,C\rangle
\leq
W_1(\mu,\nu)+2\Delta,
\]
together with a deterministic scalar certificate $U=W_1(\mu,\tilde\nu)+\Delta$ satisfying
\[
W_1(\mu,\nu)\leq U\leq W_1(\mu,\nu)+2\Delta,
\]
where $\tilde\nu=\sum_{j=1}^M(m_j/n)\delta_{z_j}$ is the collapsed measure built from the exact reconstructed cell counts.

The certificate accuracy and the realized sampler-cost bound depend only on the target-partition diameter $\Delta$, not on the field-approximation error $\eta$. The role of $\eta$ is communication: it is the quantity relevant to residual-list size, but it does not deterministically bound $\hat s$ and $s$ on its own. Bounding the residual lists from $\eta$ requires a cell-margin condition, Proposition~\ref{prop:cell-margin-vanishing-residuals}, distributional regularity, or an explicit residual-sparsity assumption. With a trivial field, the protocol degrades to the grid baseline of Proposition~\ref{prop:shared-grid-coupling}.

The compiler does not produce a cost-evaluable representation: Alice's final view determines the collapsed plan and the cell counts but not the within-cell target coordinates Bob will sample, so the realized expected sampler cost $\langle \Pi,C\rangle$ is generally not computable from her view. The scalar $U$ is what is certified, not the realized cost. If $\hat s$ is the number of occupied predicted cells and $s$ is the number of nonzero residual cells, the compiler adds
\[
\hat s\big(\lceil\log_2M\rceil+\lceil\log_2(n+1)\rceil\big)
+s\big(\lceil\log_2M\rceil+\lceil\log_2(2n+1)\rceil\big)
\]
bits beyond the preliminary field-code stage, plus $O(\log M+\log n)$ self-delimiting header bits if one insists on a prefix-free encoding of the two sparse lists. The certificate adds no communication beyond ordinary numerical precision for the collapsed OT solve.
\end{theorem}


The proof is in Appendix~\ref{app:field-code-compiler-proof}. The preliminary field-code stage may depend on both parties' inputs, and all of its bits are charged to $B_{\mathrm{field}}$. Any protocol, model, or oracle that produces a field code is completed into an exact-marginal distributed sampler with a scalar value certificate by the residual stage above; the compiler does not yield a cost-evaluable representation in the strict sense of Definition~\ref{def:certificate-bearing-output}.

\begin{proposition}[Cell-margin implies vanishing Bob residuals]
\label{prop:cell-margin-vanishing-residuals}
Suppose every matched true target lies in the interior of its public cell at distance more than $\eta$ from the cell boundary: $\operatorname{dist}(y_{\sigma(i)},\partial C_{c(i)})>\eta$ for all $i\in[n]$, where $c(i)$ is the index of the cell containing $y_{\sigma(i)}$. Then the predicted cell of $\hat y_i=\Pi_{[0,1]^d}(F_\theta(x_i))$ equals $c(i)$ for every $i$, hence $\hat m_j=m_j$ for all $j$ and Bob's residual vector is zero. In particular, the residual sparsity parameter $s$ in Theorem~\ref{thm:field-code-compiler} satisfies $s=0$.

The proposition forces $s=0$ but does not bound Alice's predicted-count support $\hat s$, which equals the number of occupied predicted cells and can still be as large as $\min(n,M)$. Therefore the displayed compiler communication is small only if $\hat s$ is also small, or if the margin condition is itself a promised subclass: on a promised margin subclass satisfying the bound above, one may run a specialized margin-certified protocol that omits Bob's residual correction and uses $\hat m=m$ directly. Without a verifiable margin promise, the residual exchange in Theorem~\ref{thm:field-code-compiler} should be retained, and the predicted-count list still costs $\hat s\big(\lceil\log_2M\rceil+\lceil\log_2(n+1)\rceil\big)$ bits.

Without the margin condition, $\eta$ alone does not bound either residual list: a target $y_k$ within distance $\le\eta$ of a cell boundary can fall on the opposite side of the boundary from $\hat y_k$, producing a residual entry. Field accuracy controls residual communication only through margin, distributional regularity, or explicit sparsity assumptions.
\end{proposition}


\subsection{Adaptive Local-Affine Transport Codes}
\label{subsec:adaptive-local-affine-transport-codes}

\begin{theorem}[Adaptive local-affine protocol]
\label{thm:adaptive-local-affine-code}
Assume Assumptions~\ref{assump:bounded-euclidean-empirical-ot}--\ref{assump:smooth-empirical-monge-realization} and the adaptive local-affine branch of Assumption~\ref{assump:field-complexity-condition}. Let $\mathcal P$ be a public dyadic partition with leaves $\mathcal L(\mathcal P)$, and let $A_Q(x)=T(c_Q)+DT(c_Q)(x-c_Q)$, $x\in Q$, be the Taylor chart at the center $c_Q$ of leaf $Q$. Suppose every leaf satisfies the Lipschitz-gradient condition $\operatorname{Lip}(DT;Q)\operatorname{diam}(Q)^2\leq 8\tau$, and a preliminary field-code stage communicates a quantized piecewise-affine field $\widetilde F_{\mathcal P}$ with $\sup_x\|\widetilde F_{\mathcal P}(x)-F_{\mathcal P}(x)\|_2\leq\lambda$, where $F_{\mathcal P}(x)=A_Q(x)$ for $x\in Q$. Then the field-preconditioned residual protocol of Theorem~\ref{thm:field-code-compiler} outputs a value-certified sampler with exact marginals and
\[
\mathbb{E}_{\mathrm{prot}}\langle \Pi_{\mathcal P},C\rangle
\leq
W_1(\mu,\nu)+2\Delta,
\qquad
W_1(\mu,\nu)\leq U\leq W_1(\mu,\nu)+2\Delta.
\]
The field accuracy parameters $\tau$ and $\lambda$ are the empirical-field-error scale $\eta=\tau+\lambda$ relevant to the residual stage of Theorem~\ref{thm:field-code-compiler}, but they do not by themselves bound $\hat s$ and $s$; a margin or sparsity condition is needed for that. They do not enter the value-certificate bound.

If each scalar affine coefficient is sent using $b$ bits, one valid field budget is
\[
B_{\mathrm{field}}(\mathcal P,b)
\leq
\sum_{Q\in\mathcal L(\mathcal P)}d\,\ell(Q)
+|\mathcal L(\mathcal P)|d(d+1)b,
\]
where $\ell(Q)$ is the dyadic depth of $Q$. See full proof in Appendix~\ref{app:adaptive-local-affine-proof}.
\end{theorem}


\begin{corollary}[Local-affine field-bit rate]
\label{cor:adaptive-local-affine-rate}
Let $K_\tau(T)$ be the minimum number of dyadic leaves needed so that every leaf satisfies the Lipschitz-gradient condition in Theorem~\ref{thm:adaptive-local-affine-code}, and let $L_\tau$ be the maximum depth of such a tree. Assume public coefficient ranges for the affine charts; if these ranges are not constant-size, the range, sign, and exponent bits are included in $b$. Choose $\Delta\leq\varepsilon/2$ so that the value certificate satisfies $W_1(\mu,\nu)\leq U\leq W_1(\mu,\nu)+\varepsilon$ by Theorem~\ref{thm:adaptive-local-affine-code}, and choose $\tau=\Theta(\varepsilon)$, $\lambda=\Theta(\varepsilon)$ with $b=O(\log(1/\varepsilon))$ bits per scalar coefficient to keep the empirical field error $\eta=\tau+\lambda$ at scale $\varepsilon$. The local-affine field code then has field-description budget
\[
O\!\left(K_\tau(T)dL_\tau+K_\tau(T)d(d+1)\log(1/\varepsilon)\right)
\]
bits, or $\widetilde O(K_\tau(T)d(d+1)+K_\tau(T)dL_\tau)$, plus the sparse residual lists from Theorem~\ref{thm:field-code-compiler}. If $T$ is $C^{1,1}$ with $\operatorname{Lip}(DT)\leq H$ globally and a uniform dyadic mesh is used, then $K_\tau(T)\leq O\!\left((H/\varepsilon)^{d/2}\right)$, recovering the same worst-case exponent as tensor-product splines while allowing smaller budgets when curvature is spatially localized. The value-certificate accuracy is controlled by $\Delta$; the choice of $\tau,\lambda$ at the same scale serves the residual-sparsity objective rather than the certificate bound. Full proof is in Appendix~\ref{app:adaptive-local-affine-rate-proof}
\end{corollary}

\subsection{Tensor-Product Spline Transport Codes}
\label{subsec:tensor-product-spline-transport-codes}

\begin{theorem}[Tensor-product spline protocol]
\label{thm:tensor-product-spline-code}
Assume Assumptions~\ref{assump:bounded-euclidean-empirical-ot}--\ref{assump:smooth-empirical-monge-realization} and the tensor-product spline branch of Assumption~\ref{assump:field-complexity-condition}. Let $G_h$ be a public tensor grid with spacing $h=1/m$, and let $\widetilde S_h$ be the multilinear spline reconstructed from quantized nodal values satisfying $\|\widetilde T(z)-T(z)\|_\infty\leq\rho$ for all $z\in G_h$. If needed, clip these nodal values coordinatewise to $[0,1]^d$ before interpolation; this cannot increase the nodal error because $T(z)\in[0,1]^d$. Then the empirical field-error parameter is
\[
\eta=\sup_x\|\widetilde S_h(x)-T(x)\|_2
\leq
\frac{d^{3/2}}{8}Hh^2+\sqrt d\,\rho,
\]
and the field-preconditioned residual protocol of Theorem~\ref{thm:field-code-compiler} outputs a value-certified sampler with exact marginals and
\[
\mathbb{E}_{\mathrm{prot}}\langle \Pi_h,C\rangle
\leq
W_1(\mu,\nu)+2\Delta,
\qquad
W_1(\mu,\nu)\leq U\leq W_1(\mu,\nu)+2\Delta.
\]
The spline interpolation bound and the nodal quantization error set the empirical-field-error scale $\eta$ relevant to the residual stage of Theorem~\ref{thm:field-code-compiler}; bounding $\hat s$ and $s$ from this $\eta$ requires a margin or sparsity assumption. They do not enter the value-certificate bound.

If every scalar control value uses $b$ bits, then the spline field uses $B_{\mathrm{field}}(h,b)=d(m+1)^d b$ bits.
\end{theorem}


The theorem gives the explicit low-dimensional field-bit law
\[
B_{\mathrm{field,spline}}(\varepsilon)=\widetilde O(\varepsilon^{-d/2})
\]
when the interpolation term dominates and quantization is chosen at the same scale. The full sampler and scalar value certificate add the residual lists from Theorem~\ref{thm:field-code-compiler}, so the total communication is
\[
B_{\mathrm{total}}(\varepsilon)
\leq
d(m+1)^d b
+\hat s\big(\lceil\log_2M\rceil+\lceil\log_2(n+1)\rceil\big)
+s\big(\lceil\log_2M\rceil+\lceil\log_2(2n+1)\rceil\big)
+O(\log M+\log n).
\]
The total rate is $\widetilde O(\varepsilon^{-d/2})$ only under an explicit residual-sparsity condition $s+\hat s=\widetilde O(\varepsilon^{-d/2})$. Without such sparsity, a uniform target partition with $\Delta=O(\varepsilon)$ requires $M=\Omega(\varepsilon^{-d})$, and the worst-case residual term contributes $\widetilde O(\min(n,\varepsilon^{-d}))$. See full proof in Appendix~\ref{app:tensor-product-spline-proof}

\begin{corollary}[Value-certified smooth field-code upper rates]
\label{cor:spline-field-bit-rate}
The local-affine and spline instantiations are scalar value-certified samplers in the sense of Definition~\ref{def:certificate-bearing-output}, where Alice outputs the scalar certificate $U$ from Theorem~\ref{thm:field-code-compiler} in addition to the sampler. Choose $\Delta\leq\varepsilon/2$ so that
\[
W_1(\mu,\nu)\leq U\leq W_1(\mu,\nu)+\varepsilon
\]
by Theorem~\ref{thm:field-code-compiler}, regardless of the spline interpolation parameters $h$ and $\rho$. Choosing $h=\Theta((\varepsilon/H)^{1/2})$ and $\rho=\Theta(\varepsilon)$ keeps the empirical field error $\eta=\tfrac{d^{3/2}}8Hh^2+\sqrt d\rho$ at scale $\varepsilon$, which is the residual-sparsity-relevant scale; this gives a field-bit budget $B_{\mathrm{field,spline}}(\varepsilon)=\widetilde O(\varepsilon^{-d/2})$. The total rate is $\widetilde O(\varepsilon^{-d/2})$ only under the explicit residual-sparsity condition $s+\hat s=\widetilde O(\varepsilon^{-d/2})$; without such sparsity, a uniform target partition forces $M=\Omega(\varepsilon^{-d})$ and the worst-case residual term contributes $\widetilde O(\min(n,\varepsilon^{-d}))$. Full proof is in Appendix~\ref{app:spline-field-bit-rate-proof}.
\end{corollary}


\subsection{Certified Lower Bounds and Sampler Separation}
\label{subsec:certified-lower-bounds-and-sampler-separation}

The lower bounds use the standard randomized public-coin communication model. They apply to all three certificate-bearing output models in Definition~\ref{def:certificate-bearing-output}: cost-evaluable, cost-certified, and scalar value-certified. The hard predicate is Gap-Hamming: under the promise $d_H(u,v)\leq m/2-\beta\sqrt m$ or $d_H(u,v)\geq m/2+\beta\sqrt m$, distinguishing the two cases requires $\Omega(m)$ communication \cite{sherstov2012ghd}.

\begin{lemma}[Certificate output implies cost estimation on the success event]
\label{lem:certificate-output-implies-cost-estimation}
If an additive-$\varepsilon$ cost-evaluable, cost-certified, or scalar value-certified protocol succeeds with probability at least $2/3$ in the sense of Definition~\ref{def:certificate-bearing-output}, then with probability at least $2/3$ Alice's final view determines a value $\widehat W$ satisfying $W_1(\mu,\nu)\leq \widehat W\leq W_1(\mu,\nu)+\varepsilon$. Hence thresholding $\widehat W$ gives a randomized communication protocol for any embedded predicate that depends on $W_1(\mu,\nu)$. Full proof is in Appendix~\ref{app:certificate-output-implies-cost-estimation-proof}.
\end{lemma}


\begin{theorem}[Bounded separated-support certified lower bound]
\label{thm:bounded-support-separated-gap-hamming}
Fix $h\in(0,\sqrt3)$ and let $L=3m+1$. For $u,v\in\{0,1\}^m$, define empirical measures in $[0,1]^2$ by
\[
\mu_u=\frac1m\sum_{i=1}^m\delta_{(3i/L,hu_i/L)},
\qquad
\nu_v=\frac1m\sum_{i=1}^m\delta_{((3i+1)/L,hv_i/L)}.
\]
Then
\[
W_1(\mu_u,\nu_v)
=
\frac1L\left(1+(\sqrt{1+h^2}-1)\frac{d_H(u,v)}{m}\right).
\]
Consequently, additive-$\varepsilon$ cost-evaluable, cost-certified, or scalar value-certified coupling output on this bounded-support family requires $\Omega(m)$ communication whenever $\varepsilon<\beta\frac{\sqrt{1+h^2}-1}{L\sqrt m}$. See Appendix~\ref{app:bounded-support-separated-gap-hamming-proof} for full proof.
\end{theorem}


Because $L=\Theta(m)$, Theorem~\ref{thm:bounded-support-separated-gap-hamming} gives the elementary certified rate $\Omega(\varepsilon^{-2/3})$. The next theorem recovers the same exponent at $d=2$ while adding a smooth diffeomorphic realization, and extends the exponent to all fixed dimensions.

\begin{theorem}[Smooth cell-packing lower bound]
\label{thm:smooth-cell-packing-diffeomorphism}
Fix $d\geq1$ and an integer $q\geq2$. Let $\phi(t)=t^3(1-t)^3\mathbf 1_{[0,1]}(t)$, $\Psi(\xi)=\prod_{r=1}^d\phi(\xi_r)$, and set $\kappa_d=\Psi(1/2,\ldots,1/2)=64^{-d}$. Let $B_{1,d}=\sup_\xi\|\nabla\Psi(\xi)\|_2$ and $B_{2,d}=\sup_\xi\|D^2\Psi(\xi)\|_{\mathrm{op}}$, and choose $0<\alpha\leq \min\{1,(4B_{1,d})^{-1}\}$. For $m=q^d$, encode $u\in\{0,1\}^{[q]^d}$ by
\[
F_u(x)=x+\alpha q^{-2}\sum_{c\in[q]^d}u_c\Psi(qx-c+\mathbf 1)e_1.
\]
At cell centers $z_c=\frac{c-\frac12\mathbf 1}{q}$, define
\[
\mu_u=\frac1m\sum_{c\in[q]^d}\delta_{F_u(z_c)},
\qquad
\nu_v=\frac1m\sum_{c\in[q]^d}\delta_{F_v(z_c)}.
\]
Then every $F_u$ is a global $C^2$ diffeomorphism, the optimal coupling is induced by $T_{u,v}=F_v\circ F_u^{-1}$, the map satisfies
\[
\|D^2T_{u,v}\|_\infty
:=
\sup_x\|D^2T_{u,v}(x)\|_{\mathrm{op}}
\leq
H_{d,\alpha}:=\frac{128}{27}\alpha B_{2,d},
\]
and
\[
W_1(\mu_u,\nu_v)
=
\frac{\alpha\kappa_d}{m^{1+2/d}}d_H(u,v).
\]
Therefore additive-$\varepsilon$ cost-evaluable, cost-certified, or scalar value-certified coupling output on this smooth $d$-dimensional family requires $\Omega(m)$ communication whenever $\varepsilon<\frac{\beta\alpha\kappa_d}{m^{1/2+2/d}}$, equivalently $\Omega(\varepsilon^{-2d/(d+4)})$ communication along the infinite sequence $m=q^d$. See proof in Appendix~\ref{app:smooth-cell-packing-diffeomorphism-proof}.
\end{theorem}


\begin{proposition}[Sampler separation for the lower-bound gadgets]
\label{prop:zero-communication-sampler-separation}
The lower-bound families in Theorems~\ref{thm:bounded-support-separated-gap-hamming} and~\ref{thm:smooth-cell-packing-diffeomorphism} do not lower-bound uncoded distributed samplers. Each admits a zero-communication exact distributed sampler. Proof is in Appendix \ref{app:zero-communication-sampler-separation-proof}.
\end{proposition}




Proposition~\ref{prop:zero-communication-sampler-separation} is the reason the paper separates sampler upper bounds from certified lower bounds. Cost-estimation hardness alone is not a sampler lower bound.

\subsection{Tightness}
\label{subsec:tightness}

The upper and lower exponents should not be read as a matched rate theorem. The spline upper bound counts field-description bits, while the lower bound is a total communication lower bound for certificate-bearing outputs. The gap reflects basis choice and residual sparsity. Appendix~\ref{app:tightness} discusses this distinction and states an open sampler-hardness problem.

\section{Empirical Study}
\label{sec:experiments}

The experiments evaluate the communication objects predicted by the theory under matched bit accounting. Each protocol summarizes the target distribution under a bit budget, and we compare the resulting certified Wasserstein-1 cost with an exact empirical-OT reference over repeated seeds. The reported metric is mean absolute relative error in the certified cost; the bit budget includes both the communicated summary and the residual cell counts from Theorem~\ref{thm:field-code-compiler}.

We compare five protocol families. \texttt{grid}\(_M\) sends public grid cell counts, matching Proposition~\ref{prop:shared-grid-coupling}; \texttt{kmeans}\(_K\) sends weighted target prototypes; and \texttt{affine\_flow}, \texttt{local\_affine}\(_K^{(b)}\), and \texttt{spline}\(_m\) communicate transport fields, corresponding respectively to a global affine map, the adaptive local-affine code of Theorem~\ref{thm:adaptive-local-affine-code}, and the tensor-product spline code of Theorem~\ref{thm:tensor-product-spline-code}. Fields are fit to paired targets on synthetic and semi-synthetic tasks, and to exact empirical-OT targets on MNIST--USPS and DOTmark. Additional details and full results are in Appendix~\ref{app:additional-empirical-details}.

\begin{figure}[!hbt]
\centering
\includegraphics[width=\linewidth]{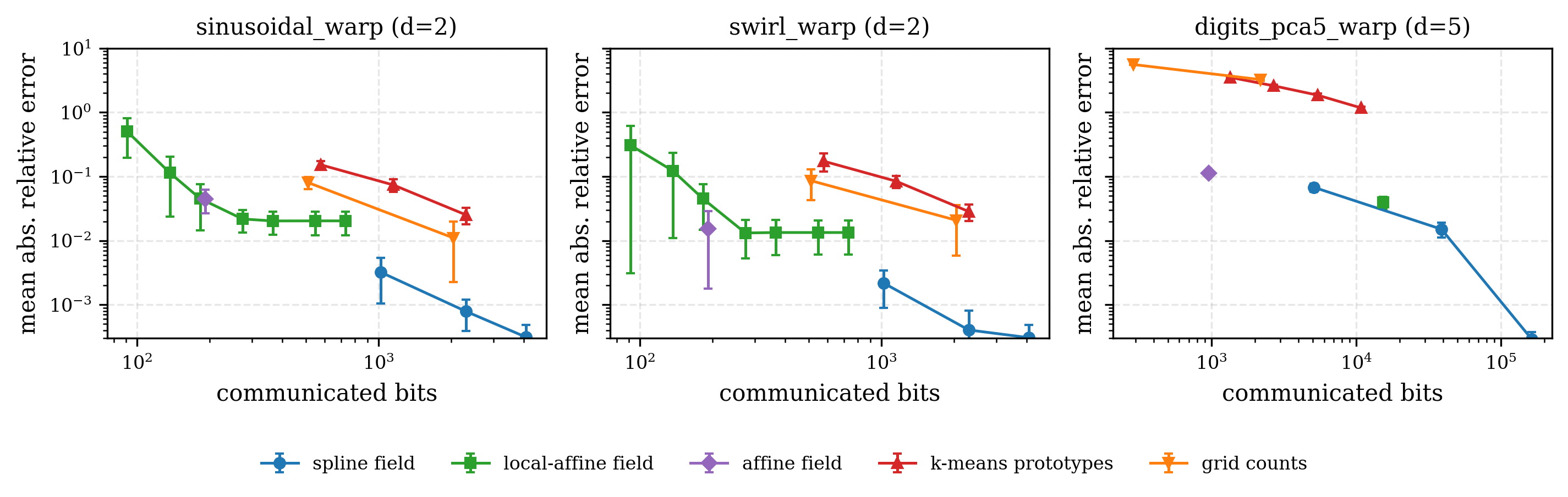}
\caption{Mean absolute relative error of the certified transport cost as a function of communicated bits on \texttt{sinusoidal\_warp} and \texttt{swirl\_warp} (\(d=2\), \(n=128\)) and on \texttt{digits\_pca5\_warp} (\(d=5\), \(n=256\)). Error bars are sample standard deviations over \(5\) seeds. The lowest plotted bit budget per family is its minimum valid encoding cost in the corresponding dimension.}
\label{fig:frontier}
\vspace{-0.2cm}
\end{figure}

The benchmarks cover smooth 2D maps, nonlinear 2D maps, 5D scale-ups, and natural unpaired transport. Figure~\ref{fig:frontier} shows the main pattern: on nonlinear 2D tasks, spline fields dominate the frontier by more than an order of magnitude over grid and \(k\)-means summaries. The local-affine sweep improves as coefficient precision \(b\) increases until the fixed dyadic curvature error dominates, matching the \(\eta=\tau+\lambda\) structure of Theorem~\ref{thm:adaptive-local-affine-code}. Refining the spline grid gives a direct multiplicative improvement, consistent with the \(h^2\) scaling of Theorem~\ref{thm:tensor-product-spline-code}.

Table~\ref{tab:scaleup-results} gives the high-dimensional comparison. Spline fields give the smallest error in every 5D setting tested; for \(n=256\), \texttt{spline\_4} reaches \(0.001672\pm0.000341\) on \texttt{gaussian\_nonlinear\_d5} and \(0.000287\pm0.000085\) on \texttt{digits\_pca5\_warp}, while grid and prototype summaries are one to three orders of magnitude worse. At \(n=1024\), field methods remain clearly separated from support summaries.

\begin{table}[!hbt]
\centering
\caption{Mean absolute relative error on the 5D scale-up benchmarks.}
\label{tab:scaleup-results}
\small
\setlength{\tabcolsep}{2.5pt}
\renewcommand{\arraystretch}{1.35}
\begin{tabular*}{\linewidth}{@{\extracolsep{\fill}}lcccccc@{}}
\toprule
Dataset & \(n\) &
\texttt{spline\_4} &
\texttt{spline\_3} &
\shortstack{\texttt{local}\\\texttt{affine\_2}} &
\texttt{kmeans\_64} &
\texttt{grid\_3} \\
\midrule
\texttt{gaussian\_nonlinear\_d5}
& 256
& \shortstack{\(1.7\mathrm{e}{-3}\)\\\(\pm 3.4\mathrm{e}{-4}\)}
& \shortstack{\(1.47\mathrm{e}{-2}\)\\\(\pm 1.64\mathrm{e}{-3}\)}
& \shortstack{\(8.15\mathrm{e}{-2}\)\\\(\pm 1.24\mathrm{e}{-2}\)}
& \shortstack{\(3.56\mathrm{e}{-1}\)\\\(\pm 1.89\mathrm{e}{-2}\)}
& \shortstack{\(1.19\)\\\(\pm 1.36\mathrm{e}{-2}\)} \\[0.45em]

\texttt{digits\_pca5\_warp}
& 256
& \shortstack{\(2.9\mathrm{e}{-4}\)\\\(\pm 8.5\mathrm{e}{-5}\)}
& \shortstack{\(1.51\mathrm{e}{-2}\)\\\(\pm 3.97\mathrm{e}{-3}\)}
& \shortstack{\(3.99\mathrm{e}{-2}\)\\\(\pm 8.50\mathrm{e}{-3}\)}
& \shortstack{\(1.18\)\\\(\pm 5.23\mathrm{e}{-2}\)}
& \shortstack{\(3.25\)\\\(\pm 1.09\mathrm{e}{-1}\)} \\[0.45em]

\texttt{gaussian\_nonlinear\_d5}
& 1024
& \shortstack{\(4.9\mathrm{e}{-3}\)\\\(\pm 3.2\mathrm{e}{-4}\)}
& \shortstack{\(3.06\mathrm{e}{-2}\)\\\(\pm 1.35\mathrm{e}{-3}\)}
& \shortstack{\(5.76\mathrm{e}{-2}\)\\\(\pm 1.72\mathrm{e}{-3}\)}
& \shortstack{\(5.23\mathrm{e}{-1}\)\\\(\pm 3.85\mathrm{e}{-3}\)}
& \shortstack{\(1.17\)\\\(\pm 1.69\mathrm{e}{-2}\)} \\[0.45em]

\texttt{digits\_pca5\_warp}
& 1024
& \shortstack{\(1.7\mathrm{e}{-3}\)\\\(\pm 4.6\mathrm{e}{-5}\)}
& \shortstack{\(3.59\mathrm{e}{-2}\)\\\(\pm 1.47\mathrm{e}{-3}\)}
& \shortstack{\(2.12\mathrm{e}{-2}\)\\\(\pm 2.83\mathrm{e}{-3}\)}
& \shortstack{\(1.63\)\\\(\pm 1.62\mathrm{e}{-2}\)}
& \shortstack{\(3.25\)\\\(\pm 2.11\mathrm{e}{-2}\)} \\
\bottomrule
\end{tabular*}
\end{table}

On natural unpaired benchmarks, spline fields also attain the best absolute accuracy on MNIST--USPS and remain competitive on DOTmark; the smaller DOTmark gap reflects that raster-image transport is not generated by a smooth diffeomorphism. Full natural-benchmark numbers and the cross-dataset plot are in Appendix~\ref{app:detailed-empirical-results}. Overall, the results support the field-first view: when transport has smooth or approximately smooth geometry, communicating a transport field is more bit-efficient than communicating only target counts or support prototypes.

\section{Discussion \& Conclusion}
\label{sec:dis_concl}

This paper studies empirical OT as a coupling-output communication problem rather than a distance-estimation problem. The main positive result is a field-code compiler: once a low-complexity transport field has been communicated, sparse residual target-cell counts restore exact marginals and yield a scalar value certificate \(W_1(\mu,\nu)\leq U\leq W_1(\mu,\nu)+2\Delta\). Adaptive local-affine and tensor-product spline codes instantiate this principle on smooth map-induced instances.

The lower bounds show that certificate-bearing outputs remain communication-hard even in smooth diffeomorphic families. At the same time, the sampler separation shows that cost-estimation hardness is not automatically sampler hardness: the optimal pair structure may be publicly sampleable while the cost remains hidden. This separates distributed sampler output from cost-evaluable, cost-certified, and value-certified output.

The empirical study supports the same field-first view. Under matched bit budgets, field summaries outperform target-count and support-prototype summaries on smooth synthetic tasks and remain competitive on natural benchmarks. The main limitation is that the upper bounds are conditional on the availability of a low-complexity field code and on residual sparsity or cell-margin structure when one wants the total communication to remain small. The compiler guarantees exact marginals and a scalar value certificate, but it does not by itself make the realized sampler cost evaluable from one party's view. The experiments also use paired or empirical-OT supervision to fit field summaries, so jointly learning such fields under communication constraints remains an important systems-level extension.


\bibliographystyle{apalike}
\bibliography{refs}

@inproceedings{yao1979distributive,
author = {Yao, Andrew Chi-Chih},
title = {Some complexity questions related to distributive computing(Preliminary Report)},
year = {1979},
isbn = {9781450374385},
publisher = {Association for Computing Machinery},
address = {New York, NY, USA},
booktitle = {Proceedings of the Eleventh Annual ACM Symposium on Theory of Computing},
pages = {209–213},
numpages = {5},
location = {Atlanta, Georgia, USA},
series = {STOC '79}
}

@article{sherstov2012ghd,
 author = {Sherstov, Alexander A.},
 title = {The Communication Complexity of Gap Hamming Distance},
 year = {2012},
 pages = {197--208},
 doi = {10.4086/toc.2012.v008a008},
 publisher = {Theory of Computing},
 journal = {Theory of Computing},
 volume = {8},
 number = {8},
}

@article{hastad2007setdisjointness,
 author = {H{\aa}stad, Johan and Wigderson, Avi},
 title = {The Randomized Communication Complexity of Set Disjointness},
 year = {2007},
 pages = {211--219},
 doi = {10.4086/toc.2007.v003a011},
 publisher = {Theory of Computing},
 journal = {Theory of Computing},
 volume = {3},
 number = {11},
}

@inproceedings{andoni2008emdhighdim,
author = {Andoni, Alexandr and Indyk, Piotr and Krauthgamer, Robert},
title = {Earth mover distance over high-dimensional spaces},
year = {2008},
publisher = {Society for Industrial and Applied Mathematics},
address = {USA},
booktitle = {Proceedings of the Nineteenth Annual ACM-SIAM Symposium on Discrete Algorithms},
pages = {343–352},
numpages = {10},
location = {San Francisco, California},
series = {SODA '08}
}

@inproceedings{andoni2009emdsketch,
author = {Andoni, Alexandr and Ba, Khanh Do and Indyk, Piotr and Woodruff, David},
title = {Efficient Sketches for Earth-Mover Distance, with Applications},
year = {2009},
isbn = {9780769538501},
publisher = {IEEE Computer Society},
address = {USA},
booktitle = {Proceedings of the 2009 50th Annual IEEE Symposium on Foundations of Computer Science},
pages = {324–330},
numpages = {7},
series = {FOCS '09}
}

@inproceedings{chen2022streamingemd,
author = {Chen, Xi and Jayaram, Rajesh and Levi, Amit and Waingarten, Erik},
title = {New streaming algorithms for high dimensional EMD and MST},
year = {2022},
isbn = {9781450392648},
publisher = {Association for Computing Machinery},
address = {New York, NY, USA},
booktitle = {Proceedings of the 54th Annual ACM SIGACT Symposium on Theory of Computing},
pages = {222–233},
numpages = {12},
location = {Rome, Italy},
series = {STOC 2022}
}

@InProceedings{andoni2016w2sketch,
  author =	{Andoni, Alexandr and Naor, Assaf and Neiman, Ofer},
  title =	{{Impossibility of Sketching of the 3D Transportation Metric with Quadratic Cost}},
  booktitle =	{43rd International Colloquium on Automata, Languages, and Programming (ICALP 2016)},
  pages =	{83:1--83:14},
  series =	{Leibniz International Proceedings in Informatics (LIPIcs)},
  ISBN =	{978-3-95977-013-2},
  ISSN =	{1868-8969},
  year =	{2016},
  volume =	{55},
  editor =	{Chatzigiannakis, Ioannis and Mitzenmacher, Michael and Rabani, Yuval and Sangiorgi, Davide},
  publisher =	{Schloss Dagstuhl -- Leibniz-Zentrum f{\"u}r Informatik},
  address =	{Dagstuhl, Germany}
}

@article{huang2015matching,
author = {Huang, Zengfeng and Radunovic, Bozidar and Vojnovic, Milan and Zhang, Qin},
title = {Communication complexity of approximate maximum matching in the message-passing model},
year = {2020},
issue_date = {Dec 2020},
publisher = {Springer-Verlag},
address = {Berlin, Heidelberg},
volume = {33},
number = {6},
issn = {0178-2770},
journal = {Distrib. Comput.},
month = dec,
pages = {515–531},
numpages = {17}
}

@InProceedings{cuturi2023sparsemaps,
  title = 	 {Monge, {B}regman and Occam: Interpretable Optimal Transport in High-Dimensions with Feature-Sparse Maps},
  author =       {Cuturi, Marco and Klein, Michal and Ablin, Pierre},
  booktitle = 	 {Proceedings of the 40th International Conference on Machine Learning},
  pages = 	 {6671--6682},
  year = 	 {2023},
  editor = 	 {Krause, Andreas and Brunskill, Emma and Cho, Kyunghyun and Engelhardt, Barbara and Sabato, Sivan and Scarlett, Jonathan},
  volume = 	 {202},
  series = 	 {Proceedings of Machine Learning Research},
  month = 	 {23--29 Jul},
  publisher =    {PMLR},
}

@inproceedings{cuturi2013sinkhorn,
author = {Cuturi, Marco},
title = {Sinkhorn distances: lightspeed computation of optimal transport},
year = {2013},
publisher = {Curran Associates Inc.},
address = {Red Hook, NY, USA},
booktitle = {Proceedings of the 27th International Conference on Neural Information Processing Systems - Volume 2},
pages = {2292–2300},
numpages = {9},
location = {Lake Tahoe, Nevada},
series = {NIPS'13}
}

@inproceedings{altschuler2017sinkhorn,
author = {Altschuler, Jason and Weed, Jonathan and Rigollet, Philippe},
title = {Near-linear time approximation algorithms for optimal transport via Sinkhorn iteration},
year = {2017},
isbn = {9781510860964},
publisher = {Curran Associates Inc.},
address = {Red Hook, NY, USA},
booktitle = {Proceedings of the 31st International Conference on Neural Information Processing Systems},
pages = {1961–1971},
numpages = {11},
location = {Long Beach, California, USA},
series = {NIPS'17}
}

@article{peyre2019computationalot,
author = {Peyr\'{e}, Gabriel and Cuturi, Marco},
title = {Computational Optimal Transport},
year = {2019},
issue_date = {Feb 2019},
publisher = {Now Publishers Inc.},
address = {Hanover, MA, USA},
volume = {11},
number = {5–6},
issn = {1935-8237},
journal = {Found. Trends Mach. Learn.},
month = feb,
pages = {355–607},
numpages = {257}
}

@InProceedings{lin2021projectionrobust,
  title = 	 { On Projection Robust Optimal Transport: Sample Complexity and Model Misspecification },
  author =       {Lin, Tianyi and Zheng, Zeyu and Chen, Elynn and Cuturi, Marco and Jordan, Michael},
  booktitle = 	 {Proceedings of The 24th International Conference on Artificial Intelligence and Statistics},
  pages = 	 {262--270},
  year = 	 {2021},
  editor = 	 {Banerjee, Arindam and Fukumizu, Kenji},
  volume = 	 {130},
  series = 	 {Proceedings of Machine Learning Research},
  month = 	 {13--15 Apr},
  publisher =    {PMLR},
}

@inproceedings{nadjahi2021fastsw,
 author = {Nadjahi, Kimia and Durmus, Alain and Jacob, Pierre E and Badeau, Roland and Simsekli, Umut},
 booktitle = {Advances in Neural Information Processing Systems},
 editor = {M. Ranzato and A. Beygelzimer and Y. Dauphin and P.S. Liang and J. Wortman Vaughan},
 pages = {12411--12424},
 publisher = {Curran Associates, Inc.},
 title = {Fast Approximation of the Sliced-Wasserstein Distance Using Concentration of Random Projections},
 volume = {34},
 year = {2021}
}

@inproceedings{nguyen2024rpsw,
author = {Nguyen, Khai and Zhang, Shujian and Le, Tam and Ho, Nhat},
title = {Sliced Wasserstein with random-path projecting directions},
year = {2024},
publisher = {JMLR.org},
booktitle = {Proceedings of the 41st International Conference on Machine Learning},
articleno = {1538},
numpages = {21},
location = {Vienna, Austria},
series = {ICML'24}
}

@inproceedings{staib2017streamingbarycenters,
author = {Staib, Matthew and Claici, Sebastian and Solomon, Justin and Jegelka, Stefanie},
title = {Parallel streaming Wasserstein barycenters},
year = {2017},
isbn = {9781510860964},
publisher = {Curran Associates Inc.},
address = {Red Hook, NY, USA},
booktitle = {Proceedings of the 31st International Conference on Neural Information Processing Systems},
pages = {2644–2655},
numpages = {12},
location = {Long Beach, California, USA},
series = {NIPS'17}
}

@inproceedings{dvurechensky2018decentralize,
 author = {Dvurechenskii, Pavel and Dvinskikh, Darina and Gasnikov, Alexander and Uribe, Cesar and Nedich, Angelia},
 booktitle = {Advances in Neural Information Processing Systems},
 editor = {S. Bengio and H. Wallach and H. Larochelle and K. Grauman and N. Cesa-Bianchi and R. Garnett},
 pages = {},
 publisher = {Curran Associates, Inc.},
 title = {Decentralize and Randomize: Faster Algorithm for Wasserstein Barycenters},
 volume = {31},
 year = {2018}
}

@INPROCEEDINGS{uribe2018distributedbarycenters,
  author={Uribe, César A. and Dvinskikh, Darina and Dvurechensky, Pavel and Gasnikov, Alexander and Nedić, Angelia},
  booktitle={2018 IEEE Conference on Decision and Control (CDC)}, 
  title={Distributed Computation of Wasserstein Barycenters Over Networks}, 
  year={2018},
  volume={},
  number={},
  pages={6544-6549},
}

@ARTICLE{cisnerosvelarde2020displacement,
  author={Cisneros-Velarde, Pedro and Bullo, Francesco},
  journal={IEEE Transactions on Control of Network Systems}, 
  title={Distributed Wasserstein Barycenters via Displacement Interpolation}, 
  year={2023},
  volume={10},
  number={2},
  pages={785-795},
}

@article{krishnan2025distributedot,
author = {Krishnan, Vishaal and Mart\'{\i}nez, Sonia},
title = {Distributed online optimization for multi-agent optimal transport},
year = {2025},
issue_date = {Jan 2025},
publisher = {Pergamon Press, Inc.},
address = {USA},
volume = {171},
number = {C},
issn = {0005-1098},
journal = {Automatica},
month = jan,
numpages = {8}
}

@article{kulcsar2025federatedsinkhorn,
  title={Federated Sinkhorn},
  author={Kulcsar, Jeremy and Kungurtsev, Vyacheslav and Korpas, Georgios and Giaconi, Giulio and Shoosmith, William},
  journal={arXiv preprint arXiv:2502.07021},
  year={2025}
}

@inproceedings{
lipman2023flowmatching,
title={Flow Matching for Generative Modeling},
author={Yaron Lipman and Ricky T. Q. Chen and Heli Ben-Hamu and Maximilian Nickel and Matthew Le},
booktitle={The Eleventh International Conference on Learning Representations },
year={2023},
}

@inproceedings{
liu2023rectifiedflow,
title={Flow Straight and Fast: Learning to Generate and Transfer Data with Rectified Flow},
author={Xingchao Liu and Chengyue Gong and qiang liu},
booktitle={The Eleventh International Conference on Learning Representations },
year={2023},
}

@article{
tong2024otcfm,
title={Improving and generalizing flow-based generative models with minibatch optimal transport},
author={Alexander Tong and Kilian FATRAS and Nikolay Malkin and Guillaume Huguet and Yanlei Zhang and Jarrid Rector-Brooks and Guy Wolf and Yoshua Bengio},
journal={Transactions on Machine Learning Research},
issn={2835-8856},
year={2024},
note={Expert Certification}
}

@InProceedings{tong2024sf2m,
  title = 	 {Simulation-Free {S}chrödinger Bridges via Score and Flow Matching},
  author =       {Tong, Alexander Y. and Malkin, Nikolay and Fatras, Kilian and Atanackovic, Lazar and Zhang, Yanlei and Huguet, Guillaume and Wolf, Guy and Bengio, Yoshua},
  booktitle = 	 {Proceedings of The 27th International Conference on Artificial Intelligence and Statistics},
  pages = 	 {1279--1287},
  year = 	 {2024},
  editor = 	 {Dasgupta, Sanjoy and Mandt, Stephan and Li, Yingzhen},
  volume = 	 {238},
  series = 	 {Proceedings of Machine Learning Research},
  month = 	 {02--04 May},
  publisher =    {PMLR},
}

@inproceedings{guo2025vrfm,
author = {Guo, Pengsheng and Schwing, Alexander G.},
title = {Variational rectified flow matching},
year = {2025},
publisher = {JMLR.org},
booktitle = {Proceedings of the 42nd International Conference on Machine Learning},
articleno = {815},
numpages = {20},
location = {Vancouver, Canada},
series = {ICML'25}
}

@inproceedings{
sidheekh2022vqflows,
title={{VQ}-Flows: Vector Quantized Local Normalizing Flows},
author={Sahil Sidheekh and Chris Barton Dock and Tushar Jain and Radu Balan and Maneesh Kumar Singh},
booktitle={The 38th Conference on Uncertainty in Artificial Intelligence},
year={2022},
}

@ARTICLE{lecun1998mnist,
  author={Lecun, Y. and Bottou, L. and Bengio, Y. and Haffner, P.},
  journal={Proceedings of the IEEE}, 
  title={Gradient-based learning applied to document recognition}, 
  year={1998},
  volume={86},
  number={11},
  pages={2278-2324},
  doi={10.1109/5.726791}}

@ARTICLE{hull1994usps,
  author={Hull, J.J.},
  journal={IEEE Transactions on Pattern Analysis and Machine Intelligence}, 
  title={A database for handwritten text recognition research}, 
  year={1994},
  volume={16},
  number={5},
  pages={550-554},
}

@ARTICLE{schrieber2017dotmark,
  author={Schrieber, Jörn and Schuhmacher, Dominic and Gottschlich, Carsten},
  journal={IEEE Access}, 
  title={DOTmark – A Benchmark for Discrete Optimal Transport}, 
  year={2017},
  volume={5},
  number={},
  pages={271-282},
}

\newpage
\appendix

\section{Full Proofs}
\label{app:full-proofs}

\subsection{Proof of Theorem~\ref{thm:field-code-compiler}}
\label{app:field-code-compiler-proof}

Let $\hat y_i=\Pi_{[0,1]^d}(F_\theta(x_i))$ and $\hat\nu=\frac1n\sum_{i=1}^n\delta_{\hat y_i}$. Because $y_{\sigma(i)}\in[0,1]^d$ and Euclidean projection onto the cube is nonexpansive, $\|\hat y_i-y_{\sigma(i)}\|_2\leq\|F_\theta(x_i)-y_{\sigma(i)}\|_2\leq\eta$ for all $i$. Let $\hat m_j$ be the number of predicted points in cell $C_j$, and let $m_j$ be the number of true target points in $C_j$. The preliminary field-code stage leaves Alice able to evaluate $F_\theta$, so she can compute $\hat m_j$. Bob can compute $m_j$ and therefore the residual $r_j=m_j-\hat m_j$.

The communication protocol is:
\begin{enumerate}
    \item A preliminary field-code stage communicates $\theta$ using $B_{\mathrm{field}}$ bits.
    \item Alice sends the sparse list of occupied predicted cell counts $(j,\hat m_j)$ as an index-sorted sequence together with a self-delimiting header for its length.
    \item Bob sends the sparse list of nonzero residuals $(j,r_j)$ as an index-sorted sequence together with a self-delimiting header for its length.
    \item Alice reconstructs all true target cell counts $m_j$ and forms $\tilde\nu=\sum_{j=1}^M\frac{m_j}{n}\delta_{z_j}$.
    \item Alice computes an optimal collapsed plan $\gamma\in\mathbb R_+^{n\times M}$ between $\mu$ and $\tilde\nu$.
    \item Alice outputs the certificate $U=\sum_{i,j}\gamma_{ij}\|x_i-z_j\|_2+\Delta=W_1(\mu,\tilde\nu)+\Delta$.
    \item The distributed sampler uses public randomness to draw a cell $J$ with probability $m_J/n$. Conditional on $J=j$, Alice samples $I$ with probability $\mathbb P[I=i\mid J=j]=\gamma_{ij}/(m_j/n)$ when $m_j>0$, and Bob samples uniformly from his target atoms in $C_j$ using a public ordering.
\end{enumerate}

The residual bit count follows from the encoding: each occupied predicted count costs one cell index and one count; each residual costs one cell index and one signed integer in $\{-n,\ldots,n\}$. The length headers contribute only $O(\log M+\log n)$ extra bits, so the total communication is the displayed sparse-pair cost plus this low-order self-delimiting overhead and $B_{\mathrm{field}}$.

The sampler has exact marginals. For the source marginal,
\[
\mathbb P[I=i]
=
\sum_{j:m_j>0}\frac{m_j}{n}\frac{\gamma_{ij}}{m_j/n}
=
\sum_j\gamma_{ij}
=\frac1n.
\]
For a target atom $y_k$ in cell $C_j$, $\mathbb P[Y=y_k]=\frac{m_j}{n}\cdot\frac1{m_j}=\frac1n$. Bob never needs to know $\gamma$; only Alice uses it to sample the source conditional on the public cell.

It remains to prove the cost. Map each true target $y_k$ to the representative $z_{c(k)}$ of its cell. Since cell diameters are at most $\Delta$, $\|y_k-z_{c(k)}\|_2\leq\Delta$, so this gives a coupling between $\nu$ and $\tilde\nu$ of cost at most $\Delta$. Hence $W_1(\nu,\tilde\nu)\leq\Delta$, and by the triangle inequality for $W_1$, $W_1(\mu,\tilde\nu)\leq W_1(\mu,\nu)+\Delta$. Since $\gamma$ is optimal for $\mu$ to $\tilde\nu$, $W_1(\mu,\tilde\nu)=\sum_{i,j}\gamma_{ij}\|x_i-z_j\|_2$.

Now bound the realized sampler cost. By the construction in step 7, the sampled pair $(I,Y)$ satisfies $Y\in C_J$ chosen uniformly among Bob's targets in $C_J$, so
\[
\mathbb E\langle \Pi,C\rangle
=
\sum_{j:m_j>0}\sum_i \gamma_{ij}\,\frac{1}{m_j}\sum_{k:y_k\in C_j}\|x_i-y_k\|_2.
\]
Triangle inequality and $\|y_k-z_j\|_2\leq\Delta$ for $y_k\in C_j$ give
\[
\frac{1}{m_j}\sum_{k:y_k\in C_j}\|x_i-y_k\|_2
\leq
\|x_i-z_j\|_2+\Delta,
\]
hence
\[
\mathbb E\langle \Pi,C\rangle
\leq
\sum_{i,j}\gamma_{ij}(\|x_i-z_j\|_2+\Delta)
=
W_1(\mu,\tilde\nu)+\Delta.
\]
Combining with the previous display,
\[
\mathbb E\langle \Pi,C\rangle
\leq
W_1(\mu,\tilde\nu)+\Delta
\leq
W_1(\mu,\nu)+2\Delta.
\]

The certificate is $U=W_1(\mu,\tilde\nu)+\Delta$. The lower bound $U\geq W_1(\mu,\nu)$ follows from $W_1(\mu,\nu)\leq W_1(\mu,\tilde\nu)+W_1(\tilde\nu,\nu)\leq U$. The upper bound $U\leq W_1(\mu,\nu)+2\Delta$ follows from $W_1(\mu,\tilde\nu)\leq W_1(\mu,\nu)+\Delta$.

The field-approximation error $\eta$ does not appear in the cost or certificate bounds. It enters only through the sparsity parameters $\hat s$ and $s$: a field that predicts targets close to their true cells reduces the number of occupied predicted cells and the number of nonzero residual cells, hence the residual-list cost.

\subsection{Proof of Theorem~\ref{thm:adaptive-local-affine-code}}
\label{app:adaptive-local-affine-proof}

Fix a leaf $Q$ and $x\in Q$. For $C^{1,1}$ maps, the integral form of Taylor's theorem gives
\[
T(x)-T(c_Q)-DT(c_Q)(x-c_Q)
=
\int_0^1\big(DT(c_Q+t(x-c_Q))-DT(c_Q)\big)(x-c_Q)\,dt.
\]
Using the operator-norm Lipschitz constant $\operatorname{Lip}(DT;Q)$ and $\|x-c_Q\|_2\leq\operatorname{diam}(Q)/2$,
\[
\|F_{\mathcal P}(x)-T(x)\|_2
\leq
\int_0^1\operatorname{Lip}(DT;Q)t\|x-c_Q\|_2^2\,dt
=
\frac12\operatorname{Lip}(DT;Q)\|x-c_Q\|_2^2
\leq
\frac18\operatorname{Lip}(DT;Q)\operatorname{diam}(Q)^2
\leq \tau.
\]
Quantization gives
\[
\|\widetilde F_{\mathcal P}(x)-T(x)\|_2
\leq
\|\widetilde F_{\mathcal P}(x)-F_{\mathcal P}(x)\|_2
+\|F_{\mathcal P}(x)-T(x)\|_2
\leq
\lambda+\tau.
\]
Evaluating at $x_i$ and using $T(x_i)=y_{\sigma(i)}$ shows that the field-approximation parameter in Theorem~\ref{thm:field-code-compiler} is $\eta=\tau+\lambda$. Theorem~\ref{thm:field-code-compiler} then yields the value-certified sampler
\[
\mathbb{E}_{\mathrm{prot}}\langle \Pi_{\mathcal P},C\rangle
\leq
W_1(\mu,\nu)+2\Delta,
\qquad
W_1(\mu,\nu)\leq U\leq W_1(\mu,\nu)+2\Delta,
\]
with $\eta=\tau+\lambda$ entering only through the residual sparsity parameters $\hat s$ and $s$ of Theorem~\ref{thm:field-code-compiler}.

The field budget consists of one dyadic address per leaf plus $d(d+1)$ scalar affine coefficients per leaf, giving
\[
B_{\mathrm{field}}(\mathcal P,b)
\leq
\sum_{Q\in\mathcal L(\mathcal P)}d\,\ell(Q)
+|\mathcal L(\mathcal P)|d(d+1)b.
\]

\subsubsection{Proof of Corollary~\ref{cor:adaptive-local-affine-rate}}
\label{app:adaptive-local-affine-rate-proof}

Take $\Delta\leq\varepsilon/2$. By Theorem~\ref{thm:adaptive-local-affine-code}, the value certificate satisfies $W_1(\mu,\nu)\leq U\leq W_1(\mu,\nu)+\varepsilon$ regardless of $\tau$ and $\lambda$. By definition, a tree with $K_\tau(T)$ leaves and maximum depth $L_\tau$ satisfies the Lipschitz-gradient condition on every leaf. Encoding the dyadic leaf addresses costs at most
\[
\sum_{Q\in\mathcal L(\mathcal P)}d\,\ell(Q)
\leq
K_\tau(T)dL_\tau
\]
bits up to delimiter overhead. The affine charts require $K_\tau(T)d(d+1)$ scalar coefficients. Quantizing coefficients with $b=O(\log(1/\varepsilon))$ bits per scalar is enough to make the induced uniform field error at most $\lambda=O(\varepsilon)$ when the coefficient ranges are bounded by the smoothness and support assumptions. If ranges are not known to be constant-size, the sign, exponent, and range-description bits are included in $b$. Thus
\[
B_{\mathrm{field}}
\leq
O\!\left(K_\tau(T)dL_\tau+K_\tau(T)d(d+1)\log(1/\varepsilon)\right),
\]
which is the stated $\widetilde O$ bound. The chosen scale $\tau,\lambda=\Theta(\varepsilon)$ keeps the empirical field error $\eta=\tau+\lambda=\Theta(\varepsilon)$, which is the residual-sparsity-relevant scale rather than the certificate-relevant scale.

If $T$ is $C^{1,1}$ with $\operatorname{Lip}(DT)\leq H$ globally, a uniform grid with side length $r$ has Euclidean diameter at most $\sqrt d\,r$ and satisfies the Lipschitz-gradient condition whenever $H d r^2\leq 8\tau$. Taking $\tau=\Theta(\varepsilon)$ gives $r=\Theta((\varepsilon/(Hd))^{1/2})$ and therefore $K_\tau(T)\leq O((Hd/\varepsilon)^{d/2})$. For fixed $d$, this is $O((H/\varepsilon)^{d/2})$, as claimed.

\subsection{Proof of Theorem~\ref{thm:tensor-product-spline-code}}
\label{app:tensor-product-spline-proof}

Let $f$ be a scalar $C^2$ function on $[0,1]^d$, and let $L_j$ be linear interpolation in coordinate $j$ on mesh spacing $h$, keeping all other coordinates fixed. The multilinear interpolant is $I_hf=L_1L_2\cdots L_df$. The telescoping identity
\[
f-I_hf=(I-L_1)f+L_1(I-L_2)f+\cdots+L_1\cdots L_{d-1}(I-L_d)f
\]
and the fact that each $L_j$ is a sup-norm contraction imply
\[
\|f-I_hf\|_\infty
\leq
\sum_{j=1}^d\|(I-L_j)f\|_\infty.
\]
For each fixed choice of the other coordinates, the one-dimensional linear interpolation remainder gives $\|(I-L_j)f\|_\infty\leq \frac{h^2}{8}\|\partial_{jj}f\|_\infty$. Applying this to component $T_k$ yields $\|T_k-S_{h,k}\|_\infty\leq \frac{dHh^2}{8}$. Taking the Euclidean norm over $k=1,\ldots,d$ gives
\[
\|T(x)-S_h(x)\|_2
\leq
\frac{d^{3/2}}{8}Hh^2.
\]

Coordinatewise clipping of the quantized nodal values to $[0,1]^d$ cannot increase the nodal error because every exact nodal value $T(z)$ already lies in the cube. The multilinear basis functions are nonnegative and sum to one on each cell, so the interpolated clipped nodal values remain in $[0,1]^d$. Therefore nodal quantization error at most $\rho$ in $\ell_\infty$ norm gives $\|\widetilde S_h(x)-S_h(x)\|_\infty\leq\rho$ and hence $\|\widetilde S_h(x)-S_h(x)\|_2\leq\sqrt d\,\rho$. Combining the two bounds,
\[
\sup_x\|\widetilde S_h(x)-T(x)\|_2
\leq
\frac{d^{3/2}}{8}Hh^2+\sqrt d\,\rho.
\]
At empirical points, this equals the field error relative to $y_{\sigma(i)}$. The empirical field-error parameter $\eta=\frac{d^{3/2}}{8}Hh^2+\sqrt d\,\rho$ is the scale relevant to residual communication in Theorem~\ref{thm:field-code-compiler}, not a deterministic bound on $\hat s$ or $s$. Theorem~\ref{thm:field-code-compiler} nevertheless yields the value-certified sampler
\[
\mathbb{E}_{\mathrm{prot}}\langle \Pi_h,C\rangle
\leq
W_1(\mu,\nu)+2\Delta,
\qquad
W_1(\mu,\nu)\leq U\leq W_1(\mu,\nu)+2\Delta,
\]
independent of $\eta$. The interpolation and quantization terms enter only through the residual lists, and bounds on those lists require a margin or sparsity assumption beyond $\eta$.

There are $(m+1)^d$ grid nodes and each carries a $d$-vector with $b$ bits per scalar, so $B_{\mathrm{field}}(h,b)=d(m+1)^d b$.

\subsubsection{Proof of Corollary~\ref{cor:spline-field-bit-rate}}
\label{app:spline-field-bit-rate-proof}

Theorem~\ref{thm:field-code-compiler} proves more than an expected-cost sampler bound: after the residual stage, Alice can compute $U=W_1(\mu,\tilde\nu)+\Delta$ from her source sample, the reconstructed collapsed target counts, and the public representatives. The sharpened proof of Theorem~\ref{thm:field-code-compiler} shows
\[
W_1(\mu,\nu)\leq U\leq W_1(\mu,\nu)+2\Delta,
\]
independent of the field-error parameter $\eta$. Choosing $\Delta\leq\varepsilon/2$ gives the value-certified guarantee. The field-bit count $d(m+1)^db$ for the spline instantiation is inherited from Theorem~\ref{thm:tensor-product-spline-code}, and the scale $h=\Theta((\varepsilon/H)^{1/2})$, $\rho=\Theta(\varepsilon)$ keeps the empirical field error $\eta=\Theta(\varepsilon)$ for the residual-sparsity analysis rather than the certificate bound. The communication is unchanged because $U$ is computed from quantities already available to Alice after the residual compiler.

\subsection{Proof of Lemma~\ref{lem:certificate-output-implies-cost-estimation}}
    \label{app:certificate-output-implies-cost-estimation-proof}

Condition on the success event $E$ from Definition~\ref{def:certificate-bearing-output}; this event has probability at least $2/3$. On $E$ the output coupling $\Pi$ is valid, so $\langle \Pi,C\rangle\geq W_1(\mu,\nu)$.

In the cost-evaluable case, let $\widehat W=\langle \Pi,C\rangle$. By definition of cost-evaluability, Alice can compute $\widehat W$ from her final view, namely input, public randomness, transcript, and output representation. On $E$, the cost-evaluable guarantee yields $\widehat W\leq W_1(\mu,\nu)+\varepsilon$.

In the cost-certified case, take $\widehat W$ to be the protocol's scalar $\widehat W$; the cost-certified guarantee on $E$ is exactly $W_1(\mu,\nu)\leq \widehat W\leq W_1(\mu,\nu)+\varepsilon$.

In the scalar value-certified case, take $\widehat W=U$, where $U$ is the protocol's scalar certificate; on $E$ the same two-sided bound holds by definition of value certification.

In all three cases, with probability at least $2/3$, Alice's final view determines a value $\widehat W$ in $[W_1(\mu,\nu),W_1(\mu,\nu)+\varepsilon]$. In a communication reduction, the parties run the OT protocol on the embedded instance and Alice thresholds this final-view value to solve the hard predicate. Outside $E$ the output may be arbitrary, but the predicate is decided correctly with probability at least $2/3$.

\subsection{Proof of Theorem~\ref{thm:bounded-support-separated-gap-hamming}}
\label{app:bounded-support-separated-gap-hamming-proof}

Let $\alpha_h=\sqrt{1+h^2}-1$. For the $i$th source-target pair,
\[
\left\|\left(\frac{3i}{L},\frac{hu_i}{L}\right)
-\left(\frac{3i+1}{L},\frac{hv_i}{L}\right)\right\|_2
=
\frac1L\left(1+\alpha_h\mathbf 1\{u_i\neq v_i\}\right).
\]
For $i\neq j$, the horizontal coordinate gap between $(3i/L,hu_i/L)$ and $((3j+1)/L,hv_j/L)$ is at least $2/L$, so every cross-index match costs at least $2/L$. Since $h<\sqrt3$, the largest same-index cost $\sqrt{1+h^2}/L$ is strictly less than $2/L$. For any non-identity permutation, every moved row uses a cross-index edge of cost at least $2/L$, while the identity edge in that same row costs at most $\sqrt{1+h^2}/L<2/L$; summing over moved rows shows that any non-identity permutation is strictly more expensive than the identity. Therefore every optimal transport plan matches index $i$ to index $i$.

The exact cost is
\[
W_1(\mu_u,\nu_v)
=
\frac1m\sum_{i=1}^m
\frac1L\big(1+\alpha_h\mathbf 1\{u_i\neq v_i\}\big)
=
\frac1L\left(1+\alpha_h\frac{d_H(u,v)}{m}\right).
\]
Under the Gap-Hamming promise, the two cases are separated by $2\beta\frac{\alpha_h}{L\sqrt m}$. An additive-$\varepsilon$ cost-evaluable, cost-certified, or scalar value-certified coupling-output protocol with $\varepsilon<\beta\frac{\alpha_h}{L\sqrt m}$ would, by Lemma~\ref{lem:certificate-output-implies-cost-estimation}, estimate the OT cost accurately enough on the success event to decide Gap-Hamming by thresholding. This contradicts the $\Omega(m)$ randomized communication lower bound for Gap-Hamming.

\subsection{Proof of Theorem~\ref{thm:smooth-cell-packing-diffeomorphism}}
\label{app:smooth-cell-packing-diffeomorphism-proof}

The scalar bump $\phi(t)=t^3(1-t)^3\mathbf 1_{[0,1]}(t)$ is $C^2$ on $\mathbb R$: at $t=0$ and $t=1$, the polynomial and its first two derivatives vanish, so the zero extension matches through second order. Hence $\Psi$ and the maps $F_u$ are $C^2$.

The supports of the translated bumps $x\mapsto \Psi(qx-c+\mathbf 1)$ are the public cells $Q_c=\prod_{r=1}^d[(c_r-1)/q,c_r/q]$, and these supports are disjoint. Write $g_u=F_u-\mathrm{id}$. On an active cell,
\[
Dg_u(x)=\alpha q^{-1}u_c e_1\nabla\Psi(qx-c+\mathbf 1)^\top.
\]
Therefore $\|Dg_u(x)\|_{\mathrm{op}}\leq \alpha q^{-1}B_{1,d}\leq \alpha B_{1,d}\leq 1/4$. Thus $g_u$ is globally $1/4$-Lipschitz on $\mathbb R^d$ and compactly supported in $[0,1]^d$. For any $y\in\mathbb R^d$, the map $x\mapsto y-g_u(x)$ is a contraction and has a unique fixed point $x_y$. This fixed point satisfies $F_u(x_y)=y$, so $F_u$ is bijective on $\mathbb R^d$. Since $DF_u$ is everywhere invertible and $F_u$ is $C^2$, $F_u$ is a global $C^2$ diffeomorphism of $\mathbb R^d$. Outside $[0,1]^d$, all bumps vanish and $F_u$ is the identity. The boundary check below shows that the restriction of this global diffeomorphism maps $[0,1]^d$ onto itself, so it is also a $C^2$ self-map of the cube.

Moreover, $F_u$ maps the cube into itself. Only the first coordinate moves, and the displacement is nonnegative. On the right boundary cell, write the local coordinate as $\xi=qx-c+\mathbf 1$ and its first coordinate as $t=\xi_1\in[0,1]$. The distance to the boundary $x_1=1$ is $(1-t)/q$, while $\alpha q^{-2}\Psi(\xi)\leq \alpha q^{-2}t^3(1-t)^3\leq \frac{1-t}{q}$ because $\alpha\leq1$ and $q\geq1$. Thus no point crosses the right boundary; all other boundaries are unchanged or moved inward.

At the cell center $z_c$, the local coordinate is $(1/2,\ldots,1/2)$, so
\[
F_u(z_c)-F_v(z_c)
=
\alpha q^{-2}(u_c-v_c)\kappa_d e_1.
\]
The displaced centers remain in $[0,1]^d$: the only moved coordinate shifts by at most $\alpha\kappa_dq^{-2}$, which is smaller than the distance $1/(2q)$ from a cell center to the nearest boundary when $q\geq2$. Thus same-cell costs are
\[
\|F_u(z_c)-F_v(z_c)\|_2
=
\alpha\kappa_d q^{-2}\mathbf 1\{u_c\neq v_c\}.
\]
For $c\neq c'$, the center distance is at least $1/q$. The displacement of any center is at most $\alpha\kappa_d q^{-2}$, hence
\[
\|F_u(z_c)-F_v(z_{c'})\|_2
\geq
\frac1q-2\alpha\kappa_d q^{-2}.
\]
Since $\alpha\leq1$, $\kappa_d\leq1/64$, and $q\geq2$, this cross-cell lower bound is strictly larger than the maximum same-cell cost. Indeed,
\[
\frac1q-2\alpha\kappa_d q^{-2}
>
\alpha\kappa_d q^{-2}
\quad\Longleftrightarrow\quad
q>3\alpha\kappa_d,
\]
which holds because $q\geq2$ and $3\alpha\kappa_d\leq3/64$. The unique optimal coupling is therefore the cellwise matching $F_u(z_c)\leftrightarrow F_v(z_c)$. It is induced by $T_{u,v}=F_v\circ F_u^{-1}$.

The exact cost follows by summing the same-cell costs:
\[
W_1(\mu_u,\nu_v)
=
\frac1m\sum_c\alpha\kappa_dq^{-2}\mathbf 1\{u_c\neq v_c\}
=
\frac{\alpha\kappa_d}{mq^2}d_H(u,v)
=
\frac{\alpha\kappa_d}{m^{1+2/d}}d_H(u,v).
\]

It remains to bound the Hessian of $T_{u,v}$. On an active cell, $\|D^2F_u(x)\|_{\mathrm{op}}\leq \alpha B_{2,d}$. From $\|DF_u-I\|_{\mathrm{op}}\leq1/4$, $\|D F_u^{-1}\|_{\mathrm{op}}\leq 4/3$. Let $G_u=F_u^{-1}$. Differentiating $F_u(G_u(y))=y$ twice gives
\[
D^2G_u(y)[h,k]
=
-DF_u(G_u(y))^{-1}
D^2F_u(G_u(y))[DG_u(y)h,DG_u(y)k].
\]
Consequently,
\[
\|D^2G_u(y)\|_{\mathrm{op}}
\leq
\frac43\cdot \alpha B_{2,d}\cdot\left(\frac43\right)^2
=
\frac{64}{27}\alpha B_{2,d}.
\]
Now $T_{u,v}=F_v\circ G_u$. The chain rule yields
\[
D^2T_{u,v}(y)[h,k]
=
D^2F_v(G_u(y))[DG_u(y)h,DG_u(y)k]
+DF_v(G_u(y))D^2G_u(y)[h,k].
\]
Using $\|D^2F_v\|_{\mathrm{op}}\leq\alpha B_{2,d}$, $\|DG_u\|_{\mathrm{op}}\leq4/3$, and $\|DF_v\|_{\mathrm{op}}\leq5/4$, we get
\[
\|D^2T_{u,v}\|_\infty
\leq
\alpha B_{2,d}\left(\frac43\right)^2
+\frac54\cdot\frac{64}{27}\alpha B_{2,d}
=
\frac{128}{27}\alpha B_{2,d}.
\]

Under the Gap-Hamming promise, the exact cost identity creates a gap $2\beta\frac{\alpha\kappa_d}{m^{1/2+2/d}}$. If $\varepsilon<\frac{\beta\alpha\kappa_d}{m^{1/2+2/d}}$, Lemma~\ref{lem:certificate-output-implies-cost-estimation} would turn an additive-$\varepsilon$ cost-evaluable, cost-certified, or scalar value-certified coupling-output protocol into a Gap-Hamming protocol, requiring $\Omega(m)$ bits. Since $\varepsilon\asymp m^{-1/2-2/d}$ is equivalent to $m\asymp \varepsilon^{-2d/(d+4)}$, the $\varepsilon$-dependent lower bound is $\Omega(\varepsilon^{-2d/(d+4)})$.

\subsection{Proof of Proposition~\ref{prop:zero-communication-sampler-separation}}
\label{app:zero-communication-sampler-separation-proof}

For Theorem~\ref{thm:bounded-support-separated-gap-hamming}, the public randomness samples $i\in[m]$ uniformly. Alice outputs her $i$th source atom $\left(\frac{3i}{L},\frac{hu_i}{L}\right)$, and Bob outputs his $i$th target atom $\left(\frac{3i+1}{L},\frac{hv_i}{L}\right)$. The proof of Theorem~\ref{thm:bounded-support-separated-gap-hamming} shows that this indexwise matching is optimal for every $u,v$. The protocol communicates zero bits and produces the exact optimal distributed sampler. The cost still depends on $d_H(u,v)$, so a cost certificate is not produced.

For Theorem~\ref{thm:smooth-cell-packing-diffeomorphism}, the public randomness samples a cell $c\in[q]^d$ uniformly. Because $\alpha\kappa_d q^{-2}<1/(2q)$, every displaced center remains inside its original public cell, so Alice and Bob can recover the common label $c$ from their local samples without communication. Alice outputs $F_u(z_c)$ and Bob outputs $F_v(z_c)$. The proof of Theorem~\ref{thm:smooth-cell-packing-diffeomorphism} shows that the cellwise matching is optimal for every $u,v$. Again the protocol communicates zero bits and produces the exact optimal distributed sampler, while the cost remains hidden in the mismatch pattern.

\section{Tightness}
\label{app:tightness}
Corollary~\ref{cor:spline-field-bit-rate} gives a field-description upper rate $B_{\mathrm{field,spline}}(\varepsilon)=\widetilde O(\varepsilon^{-d/2})$, and Theorem~\ref{thm:smooth-cell-packing-diffeomorphism} gives a total certified-output lower rate $B_{\mathrm{cert}}(\varepsilon)=\Omega(\varepsilon^{-2d/(d+4)})$. At $d=2$, the exponents are $1$ and $2/3$. At $d=5$, they are $5/2$ and $10/9$. The two quantities are not the same complexity measure: the upper rate counts only the spline-field bits, while the lower rate is a total communication lower bound for any cost-evaluable, cost-certified, or value-certified protocol. The corresponding total value-certified upper rate is the field cost plus the residual-count lists in Theorem~\ref{thm:field-code-compiler}, and it matches the field rate $\widetilde O(\varepsilon^{-d/2})$ only on subclasses with $s+\hat s=\widetilde O(\varepsilon^{-d/2})$. Hence the displayed exponent gap is a basis-and-residual-sparsity gap rather than a matched upper-versus-lower theorem in the same total communication measure. The spline upper theorem pays for a uniform tensor basis on the field, while the lower bound creates a certified cost gap from small independent perturbations of size $q^{-2}$ in public cells. Proposition~\ref{prop:zero-communication-sampler-separation} shows that the lower-bound perturbations leave the optimal pair structure publicly determined, so they impose no sampler communication. Closing the exponent gap will require either residual-sparse upper-bound classes or a lower bound stated in the same field-complexity model.

The open tightness problem is basis-aware. The algorithmic direction asks for an adaptive field code that spends bits only on cells where the transport field has nonzero perturbation, improving over the uniform tensor rate on localized families. The converse direction asks for a sampler lower bound that makes the near-optimal pair structure itself depend on both inputs, rather than only on the certified cost value. We conjecture that adaptive certified field codes match the cell-packing exponent $\varepsilon^{-2d/(d+4)}$ on localized smooth families, and that the right complexity parameter for smooth sampler output is the metric entropy of the transport field in the communicated basis plus the residual exact-marginal correction cost.

\begin{openproblem}[Sampler-hard smooth transport]
\label{op:sampler-hard-smooth-transport}
Replace the public one-atom-per-cell gadgets by two-by-two switch cells. In each cell, Alice's private bit and Bob's private bit would jointly decide whether the lower-cost local matching is diagonal or off-diagonal, so public randomness could no longer choose the correct pair label without communication. A smooth version would pack these switch cells into separated boxes and realize their perturbations by local diffeomorphisms. The missing proof obligation is stability: cross-cell transport must remain dominated while each local switch keeps a robust cost gap. This is the next sampler lower-bound target.
\end{openproblem}

\section{Additional Empirical Details}
\label{app:additional-empirical-details}

\subsection{Benchmark Suite}
\label{app:benchmark-suite}

\begin{table}[!hbt]
\centering
\small
\setlength{\tabcolsep}{4pt}
\caption{Benchmark groups used in the empirical study.}
\label{tab:benchmark-suite}
\begin{tabularx}{\linewidth}{@{}lYp{0.22\linewidth}Y@{}}
\toprule
Group & Datasets & Sample size & Role \\
\midrule
Smooth 2D &
\shortstack[l]{\texttt{gaussian\_shift}\\ \texttt{gaussian\_affine}} &
\(n=128\), 5 seeds &
Affine-field sanity \\
Nonlinear 2D &
\shortstack[l]{\texttt{sinusoidal\_warp}\\ \texttt{swirl\_warp}} &
\(n=128\), 5 seeds &
Spline/local-field test \\
5D scale-up &
\shortstack[l]{\texttt{gaussian\_nonlinear\_d5}\\ \texttt{digits\_pca5\_warp}} &
\shortstack[l]{\(n=256\), 5 seeds;\\ \(n=1024\), 3 seeds} &
Dimension and sample-size check \\
Natural unpaired &
\shortstack[l]{\texttt{mnist\_usps\_pca5}\\ \texttt{dotmark\_classicimages\_2d}} &
\(n=512\), 5 seeds &
Real-data transport geometry \\
\bottomrule
\end{tabularx}
\end{table}
\texttt{mnist\_usps\_pca5} uses MNIST \citep{lecun1998mnist} resized from \(28\times28\) to \(16\times16\), USPS \citep{hull1994usps} normalized to \([0,1]\), and PCA-5 fit on the union so that both parties share coordinates. \texttt{dotmark\_classicimages\_2d} uses DOTmark ClassicImages \citep{schrieber2017dotmark} at resolution \(64\), treating each image as a discrete mass distribution and sampling point clouds with within-pixel jitter. In all ten real-benchmark trials, the exact empirical OT plan has \(512\) nonzero entries for \(n=512\), so the oracle target is permutation-valued.

\subsection{Additional Results}
\label{app:detailed-empirical-results}

\begin{table}[!hbt]
\centering
\caption{Mean absolute relative error on natural unpaired benchmarks.}
\label{tab:natural-unpaired-results}
\resizebox{\linewidth}{!}{
\begin{tabular}{lccccc}
\toprule
Benchmark & \texttt{spline\_4} & \texttt{spline\_3} & \texttt{local\_affine\_2} & \texttt{kmeans\_64} & \texttt{grid\_3} \\
\midrule
\texttt{mnist\_usps\_pca5} &
\(0.020219\pm0.001204\) &
\(0.028262\pm0.001418\) &
\(0.037872\pm0.001950\) &
\(0.031616\pm0.002307\) &
\(0.096345\pm0.013789\) \\
\texttt{dotmark\_classicimages\_2d} &
\(0.151760\pm0.096331\) &
\(0.201435\pm0.151991\) &
\(0.162634\pm0.123231\) &
\(0.166822\pm0.100585\) &
\(0.982010\pm0.512256\) \\
\bottomrule
\end{tabular}
}
\end{table}

\begin{figure}[!hbt]
\centering
\includegraphics[width=\linewidth]{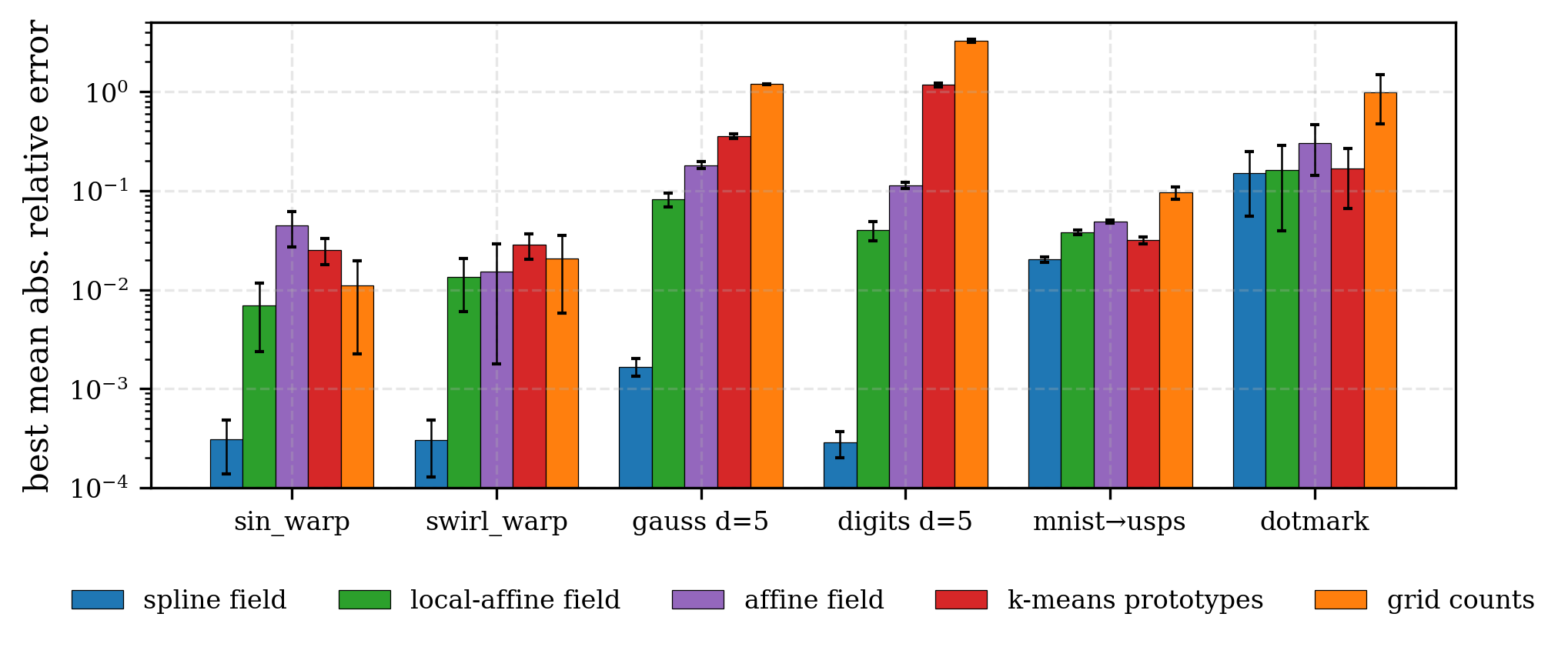}
\caption{Best per-family mean absolute relative error on each benchmark, with \(5\) seeds per benchmark and error bars sampling standard deviation; sample sizes are \(n=128\) for the 2D tasks, \(n=256\) for the 5D scale-up tasks, and \(n=512\) for the natural benchmarks.}
\label{fig:cross-dataset}
\end{figure}

On the smooth 2D tasks \texttt{gaussian\_shift} and \texttt{gaussian\_affine}, the global affine field already saturates the achievable accuracy. The frontier AUCs, seedwise mean \(\pm\) sample standard deviation, are \(0.000243\pm0.000131\) and \(0.002156\pm0.000956\) for the field family, against \(0.001847\pm0.000453\) and \(0.004388\pm0.001516\) for \(k\)-means and \(0.004337\pm0.003855\) and \(0.006057\pm0.007196\) for grid summaries. Representative individual points include \texttt{affine\_flow} at \(1.51\cdot10^{-4}\pm1.08\cdot10^{-4}\) on \texttt{gaussian\_shift} and \(4.46\cdot10^{-4}\pm5.05\cdot10^{-4}\) on \texttt{gaussian\_affine}, both at \(160\) bits.

On the nonlinear 2D tasks \texttt{sinusoidal\_warp} and \texttt{swirl\_warp}, spline fields dominate the frontier, with AUCs \(0.002716\pm0.001307\) and \(0.001673\pm0.000605\), by more than an order of magnitude over local-affine \((0.037643,0.037393)\), affine \((0.044503,0.015347)\), grid \((0.091259,0.106692)\), and \(k\)-means \((0.162777,0.185490)\). The strongest spline points reach \(3.11\cdot10^{-4}\pm1.74\cdot10^{-4}\) on \texttt{sinusoidal\_warp} and \(3.06\cdot10^{-4}\pm1.76\cdot10^{-4}\) on \texttt{swirl\_warp}. On \texttt{swirl\_warp}, the local-affine error decreases from \(0.309\pm0.437\) at \(b=4\), \(91.2\) bits, to a plateau of \(0.0134\pm0.007\) at \(b\geq12\).

On natural unpaired benchmarks, spline fields again attain the best absolute accuracy. On \texttt{mnist\_usps\_pca5}, \texttt{spline\_4} reaches \(0.020219\pm0.001204\) against \(0.031616\pm0.002307\) for \texttt{kmeans\_64} and \(0.096345\pm0.013789\) for \texttt{grid\_3}. On \texttt{dotmark\_classicimages\_2d}, the spline error is \(0.151760\pm0.096331\) versus \(0.166822\pm0.100585\) for \texttt{kmeans\_64} and \(0.982010\pm0.512256\) for \texttt{grid\_3}; the closer race reflects that the underlying transport between raster images is not generated by a smooth diffeomorphism.

\newpage
\section*{NeurIPS Paper Checklist}

The checklist is designed to encourage best practices for responsible machine learning research, addressing issues of reproducibility, transparency, research ethics, and societal impact. Do not remove the checklist: {\bf The papers not including the checklist will be desk rejected.} The checklist should follow the references and follow the (optional) supplemental material.  The checklist does NOT count towards the page
limit. 

Please read the checklist guidelines carefully for information on how to answer these questions. For each question in the checklist:
\begin{itemize}
    \item You should answer \answerYes{}, \answerNo{}, or \answerNA{}.
    \item \answerNA{} means either that the question is Not Applicable for that particular paper or the relevant information is Not Available.
    \item Please provide a short (1–2 sentence) justification right after your answer (even for NA). 
\end{itemize}

{\bf The checklist answers are an integral part of your paper submission.} They are visible to the reviewers, area chairs, senior area chairs, and ethics reviewers. You will be asked to also include it (after eventual revisions) with the final version of your paper, and its final version will be published with the paper.

The reviewers of your paper will be asked to use the checklist as one of the factors in their evaluation. While "\answerYes{}" is generally preferable to "\answerNo{}", it is perfectly acceptable to answer "\answerNo{}" provided a proper justification is given (e.g., "error bars are not reported because it would be too computationally expensive" or "we were unable to find the license for the dataset we used"). In general, answering "\answerNo{}" or "\answerNA{}" is not grounds for rejection. While the questions are phrased in a binary way, we acknowledge that the true answer is often more nuanced, so please just use your best judgment and write a justification to elaborate. All supporting evidence can appear either in the main paper or the supplemental material, provided in appendix. If you answer \answerYes{} to a question, in the justification please point to the section(s) where related material for the question can be found.

IMPORTANT, please:
\begin{itemize}
    \item {\bf Delete this instruction block, but keep the section heading ``NeurIPS Paper Checklist"},
    \item  {\bf Keep the checklist subsection headings, questions/answers and guidelines below.}
    \item {\bf Do not modify the questions and only use the provided macros for your answers}.
\end{itemize}


\begin{enumerate}

\item {\bf Claims}
    \item[] Question: Do the main claims made in the abstract and introduction accurately reflect the paper's contributions and scope?
    \item[] Answer: \answerYes{}.
    \item[] Justification: The abstract and introduction state the paper's scope as empirical OT coupling-output communication, distinguish the output models, and summarize the field-code compiler, smooth certified lower bounds, sampler/certificate separation, and empirical bit-budget comparisons. The claims are matched by the formal results in Sections~\ref{sec:problem-formulation}--\ref{sec:theoretical-results}, the experiments in Section~\ref{sec:experiments}, and the limitations in Section~\ref{sec:dis_concl}.

\item {\bf Limitations}
    \item[] Question: Does the paper discuss the limitations of the work performed by the authors?
    \item[] Answer: \answerYes{}.
    \item[] Justification: Section~\ref{sec:discussion-conclusion} discusses the main limitations: the upper bounds require a low-complexity field code, total communication depends on residual sparsity or cell-margin structure, the compiler gives a scalar value certificate but not a cost-evaluable representation, and the experiments fit fields using paired or empirical-OT supervision.

\item {\bf Theory assumptions and proofs}
    \item[] Question: For each theoretical result, does the paper provide the full set of assumptions and a complete (and correct) proof?
    \item[] Answer: \answerYes{}.
    \item[] Justification: The standing assumptions are stated in Section~\ref{subsec:standing-assumptions}, and the theorem, proposition, corollary, and lemma statements reference the relevant assumptions and output models. Full proofs are provided in Appendix~\ref{app:full-proofs}, with the main results cross-referenced from Section~\ref{sec:theoretical-results}.

\item {\bf Experimental result reproducibility}
    \item[] Question: Does the paper fully disclose all the information needed to reproduce the main experimental results of the paper to the extent that it affects the main claims and/or conclusions of the paper (regardless of whether the code and data are provided or not)?
    \item[] Answer: \answerYes{}.
    \item[] Justification: Section~\ref{sec:experiments} and Appendix~\ref{app:additional-empirical-details} describe the protocol families, benchmark groups, sample sizes, seeds, metrics, bit accounting, and how field summaries are fit. The experiments use synthetic or publicly available datasets, and the reported tables and figures specify the main comparisons needed to verify the empirical claims.

\item {\bf Open access to data and code}
    \item[] Question: Does the paper provide open access to the data and code, with sufficient instructions to faithfully reproduce the main experimental results, as described in supplemental material?
    \item[] Answer: \answerNo{}.
    \item[] Justification: The paper describes the synthetic generation procedure, public datasets, protocols, and evaluation setup, but the source code is not openly released at submission time. The source code will be released after acceptance with scripts and instructions to reproduce the main experimental results.

\item {\bf Experimental setting/details}
    \item[] Question: Does the paper specify all the training and test details (e.g., data splits, hyperparameters, how they were chosen, type of optimizer, etc.) necessary to understand the results?
    \item[] Answer: \answerYes{}.
    \item[] Justification: Section~\ref{sec:experiments} and Appendix~\ref{app:additional-empirical-details} specify the protocol families, benchmark datasets, sample sizes, number of seeds, bit accounting, field-fitting supervision, and evaluation metric. The experiments do not train neural networks; the main algorithmic choices are the communicated summaries and their resolutions or coefficient precisions.

\item {\bf Experiment statistical significance}
    \item[] Question: Does the paper report error bars suitably and correctly defined or other appropriate information about the statistical significance of the experiments?
    \item[] Answer: \answerYes{}.
    \item[] Justification: The main figure and result tables report seedwise mean \(\pm\) sample standard deviation over repeated seeds. Section~\ref{sec:experiments} and Appendix~\ref{app:detailed-empirical-results} state the number of seeds used for each benchmark and identify the error bars as sample standard deviations.

\item {\bf Experiments compute resources}
    \item[] Question: For each experiment, does the paper provide sufficient information on the computer resources (type of compute workers, memory, time of execution) needed to reproduce the experiments?
    \item[] Answer: \answerNo{}.
    \item[] Justification: The current version reports the experimental setup and repeated-seed evaluation but does not yet provide detailed compute-worker type, memory, runtime per run, or total compute. These details will be added to the supplemental material together with the code release after acceptance.
    
\item {\bf Code of ethics}
    \item[] Question: Does the research conducted in the paper conform, in every respect, with the NeurIPS Code of Ethics \url{https://neurips.cc/public/EthicsGuidelines}?
    \item[] Answer: \answerYes{}.
    \item[] Justification: The paper is theoretical and empirical work on optimal transport communication, uses synthetic and publicly available benchmark datasets, and does not involve human-subject experiments, private data collection, or deployment of a high-risk model. The work is intended to conform to the NeurIPS Code of Ethics.

\item {\bf Broader impacts}
    \item[] Question: Does the paper discuss both potential positive societal impacts and negative societal impacts of the work performed?
    \item[] Answer: \answerYes{}.
    \item[] Justification: The paper discusses the foundational nature of the work and its potential positive impact on communication-efficient distributed OT and flow training. It also notes limitations and risks relevant to distributed settings, including that communicating summaries or fields does not by itself provide privacy guarantees.
    
\item {\bf Safeguards}
    \item[] Question: Does the paper describe safeguards that have been put in place for responsible release of data or models that have a high risk for misuse (e.g., pretrained language models, image generators, or scraped datasets)?
    \item[] Answer: \answerNA{}.
    \item[] Justification: The paper does not release pretrained language models, image generators, scraped datasets, or other high-risk assets. The experiments use synthetic data and standard public benchmark datasets.

\item {\bf Licenses for existing assets}
    \item[] Question: Are the creators or original owners of assets (e.g., code, data, models), used in the paper, properly credited and are the license and terms of use explicitly mentioned and properly respected?
    \item[] Answer: \answerYes{}.
    \item[] Justification: The paper cites the original sources for the public datasets used in the empirical study, including MNIST, USPS, and DOTmark. The datasets are used only as standard research benchmarks; no new redistribution of these datasets is introduced by the paper.

\item {\bf New assets}
    \item[] Question: Are new assets introduced in the paper well documented and is the documentation provided alongside the assets?
    \item[] Answer: \answerNA{}.
    \item[] Justification: The paper does not introduce a new dataset, pretrained model, or benchmark asset. The planned code release after acceptance will document the synthetic data generation and experimental scripts.

\item {\bf Crowdsourcing and research with human subjects}
    \item[] Question: For crowdsourcing experiments and research with human subjects, does the paper include the full text of instructions given to participants and screenshots, if applicable, as well as details about compensation (if any)? 
    \item[] Answer: \answerNA{}.
    \item[] Justification: The paper does not involve crowdsourcing, user studies, or research with human subjects. All experiments use synthetic data or standard public benchmark datasets.

\item {\bf Institutional review board (IRB) approvals or equivalent for research with human subjects}
    \item[] Question: Does the paper describe potential risks incurred by study participants, whether such risks were disclosed to the subjects, and whether Institutional Review Board (IRB) approvals (or an equivalent approval/review based on the requirements of your country or institution) were obtained?
    \item[] Answer: \answerNA{}.
    \item[] Justification: The paper does not involve crowdsourcing, user studies, or research with human subjects, so IRB approval or equivalent human-subject review is not applicable.

\item {\bf Declaration of LLM usage}
    \item[] Question: Does the paper describe the usage of LLMs if it is an important, original, or non-standard component of the core methods in this research? Note that if the LLM is used only for writing, editing, or formatting purposes and does not impact the core methodology, scientific rigorousness, or originality of the research, declaration is not required.
    \item[] Answer: \answerNA{}.
    \item[] Justification: LLMs are not used as part of the core methodology, theory, experiments, or scientific contribution. Any language editing or formatting assistance does not affect the originality, rigor, or results of the research.

\end{enumerate}

\end{document}